\documentclass[letterpaper,twocolumn,showpacs,nopreprintnumbers,nofootinbib,prd]{revtex4}
\usepackage[latin1]{inputenc}
\pdfoutput=1
\usepackage{amsmath}
\usepackage{amssymb}
\usepackage{amsfonts}
\usepackage{array}
\usepackage{multirow}
\usepackage{booktabs}
\usepackage{hyperref}
\usepackage[all]{hypcap}
\usepackage{graphicx} 
\usepackage{dcolumn} 
\usepackage{paralist}

\newif\ifonecol\onecolfalse

\def \diff[#1]#2 {\frac{\mathrm{d}#1}{ \mathrm{d}#2}}
\def \ind {\mathrm{d}}

\begin{document}

\title{Ultra-high neutrino fluxes as a probe for non-standard physics}
\date{\today}

\author{Atri Bhattacharya}
\email{atri@hri.res.in}
\affiliation{Harish-Chandra Research Institute, Chhatnag Road, Jhunsi, 
Allahabad 211 019, India}

\author{Sandhya Choubey}
\email{sandhya@hri.res.in}
\affiliation{Harish-Chandra Research Institute, Chhatnag Road, Jhunsi,
Allahabad 211 019, India}

\author{Raj Gandhi}
\email{nubarnu@gmail.com}
\affiliation{Harish-Chandra Research Institute, Chhatnag Road, Jhunsi,
Allahabad 211 019, India}

\author{Atsushi Watanabe}
\email{watanabe@muse.sc.niigata-u.ac.jp}
\affiliation{Department of Physics, Niigata University, Niigata 950-2181, Japan}

\begin{abstract}
We examine how light neutrinos coming from distant active galactic nuclei (AGN) and similar high energy sources may be used as tools to probe non-standard physics. In particular we discuss how studying the energy spectra of each neutrino flavour coming from such distant sources and their distortion relative to each other may serve as pointers to exotic physics such as neutrino decay, Lorentz symmetry violation, pseudo-Dirac effects, CP and CPT violation and quantum decoherence. This allows us to probe hitherto unexplored ranges of parameters for the above cases, for example lifetimes in the range $ 10^{-3}-10^{4} $ s/eV for the case of neutrino decay. We show that standard neutrino oscillations ensure that the different flavours arrive at the earth with similar shapes even if their flavour spectra at source may differ strongly in both shape and magnitude. As a result, observed differences between the spectra of various flavours at the detector would be signatures of non-standard physics altering neutrino fluxes during propagation rather than those arising during their production at source. Since detection of ultra-high energy (UHE) neutrinos is perhaps imminent, it is possible that such differences in spectral shapes will be tested in neutrino detectors in the near future. To that end, using the IceCube detector as an example, we show how our results translate to observable shower and muon-track event rates.
\end{abstract}

\pacs{14.60.Pq,14.60.Lm,14.60.St,13.15.+g,11.30.Cp,98.54.Cm,95.85.Ry}

\maketitle

\section{\label{sec:intro}Introduction}

The extraordinary success of neutrino experiments in the last 
few decades has propelled neutrino physics to the centre-stage of 
particle physics. A series of seminal observations 
\cite{Cleveland:1998nv, Abdurashitov:2002nt,Hampel:1998xg,Fukuda:2002uc,Ahmad:2002jz, Aharmim:2005gt, Aharmim:2008kc, Arpesella:2008mt, Ashie:2005ik,:2008ee, Aliu:2004sq, Adamson:2008zt} 
have 
provided us with a ``new standard model'' (nuSM), in which the standard 
model of elementary particles is augmented by three massive 
neutrinos which mix. Therefore, in addition to the standard model 
parameters, the nuSM also includes at least 2 mass squared 
differences\footnote{The mass squared differences are defined 
as $\Delta m_{ij}^2 = m_i^2 - m_j^2$.}
$\Delta m_{21}^2$ and $\Delta m_{31}^2$, three mixing angles 
$\theta_{12}$, $\theta_{23}$ and $\theta_{13}$, one (so-called) 
Dirac CP phase, and two Majorana CP phases (if neutrinos are Majorana 
particles).  
With the existence of neutrino masses and mixing 
confirmed, focus has now shifted to the next level, {\it viz.}, 
(i) making precise measurements of the known oscillation 
parameters, and (ii) 
determining the hitherto unknown properties of neutrinos. 
Among the neutrino parameters which belong to the standard nuSM 
picture and which are still unknown are the mixing angle $\theta_{13}$, 
the sign of $\Delta m_{31}^2$, and the CP phase(s). Next-generation 
neutrino experiments are expected to throw light on some or all of 
these standard neutrino oscillation parameters. Data from these experiments can also be used to probe physics beyond the standard 
paradigm. This includes a variety of new physics scenarios such as 
non-standard interactions at the source and detector as well as 
during the propagation of neutrinos \cite{Wolfenstein:1977ue,Valle:1987gv,Guzzo:1991hi}, 
non-unitarity of the neutrino mixing matrix \cite{Langacker:1988ur,Bilenky:1992wv,Bekman:2002zk}, 
violation of the equivalence principle \cite{Gasperini:1988zf,Gasperini:1989rt}, 
neutrino decay \cite{Bahcall:1972my,Beacom:2002vi,Beacom:2002cb}, 
violation of Lorentz invariance \cite{Kostelecky:2003cr,Kostelecky:2003fs}, 
pseudo-Dirac neutrinos \cite{Petcov1982245,PhysRevLett.67.1685,PhysRevD.64.013003}, 
and neutrino decoherence \cite{PhysRevLett.85.1166}. 
Among the forthcoming sets of experiments which hold promise 
for new physics searches are neutrino telescopes, which  
have been designed to observe ultra high energy neutrinos coming from 
astrophysical sources. 

Very high energy cosmic rays with energies as high as $ 10^{10} $ GeV have been 
observed \cite{Aharonian:2009xn,Acciari:2009rs}. There is also now a large body of evidence for 
high energy gamma rays coming from astrophysical sources. 
Understanding the origin and source of these high energy cosmic 
rays remains a challenge. A plethora of papers have appeared 
in the past trying to provide a viable 
model for these observations. Nearly all such models allow acceleration 
of protons to energies in the realm of $ 10^{10}-10^{11} $ GeV. Such high energy 
protons will invariably lead to the production of highly accelerated 
pions (and kaons) through $p\gamma$ and $pp$ collisions. These pions 
would in turn produce neutrinos carrying energy anywhere in the range 
of $ 10^4-10^{10} $ GeV depending on the type of source. Detectors such as 
AMANDA \cite{Baret:2008zz}, IceCube \cite{Halzen:2009tz}, BAIKAL
\cite{Aynutdinov:2009zz}, ANTARES \cite{Margiotta:2009zz}, 
KM3NET \cite{Margiotta:2010zz}, 
RICE \cite{1742-6596-81-1-012008} and ANITA \cite{1742-6596-136-2-022052} 
have been constructed (or are under construction) using techniques 
that would make it possible for them to observe these ultra high energy 
neutrinos. 

Neutrinos produced via decay of pions are expected to 
roughly carry the flavor ratio 
$ (\nu_e:\nu_\mu:\nu_\tau =)\: 1:2:0 $ at the source. 
Standard neutrino oscillations in vacuum massage this ratio 
during propagation 
to $1:1:1$ \cite{Learned:1994wg,Athar:2000yw} 
at the detector, if we assume $\theta_{13}=0$ and 
$\theta_{23}=\pi/4$ consistent with current data 
\cite{GonzalezGarcia:2010er,Bandyopadhyay:2008va,Fogli:2008jx}. 
It has also been recently stressed 
\cite{Bhattacharya:2009tx} that standard flavor oscillations over Mega-parsec distances  
make the neutrino spectra of every flavor 
nearly identical in shape. 
Therefore, if for any reason the astrophysics in the source leads to a ratio different from $ 1:2:0 $ or spectral shapes for flavours which differ widely from each other, standard oscillations still massage them into identical shapes and magnitudes which are within a factor of roughly 2 of each other by the time they reach the earth \cite{Bhattacharya:2009tx}. 

The potential of the neutrino telescopes to probe new physics 
using absolute flux ratios has been studied in 
\cite{Beacom:2002vi,Beacom:2003nh,Barenboim:2003jm,Keranen:2003xd,
Beacom:2003eu,Hooper:2004xr,Hooper:2005jp,Meloni:2006gv,Xing:2008fg,
Esmaili:2009dz}. 
In an earlier paper \cite{Bhattacharya:2009tx}, we considered 
two specific new physics scenarios, {\it viz.}, neutrino 
decay and Lorentz invariance violation, and  
showed how they affect the diffuse ultra high energy neutrino flux. 
We emphasised how 
spectral information could be used 
to extract new physics from the ultra high energy neutrino data. 
In this paper we extend our earlier analysis to 
include more new physics cases. In addition to 
neutrino decay and Lorentz invariance violation we consider the effect of 
pseudo-Dirac neutrinos and neutrino decoherence during propagation. Further, we include three flavour effects, by allowing the mixing angle $ \sin^2\left( \theta_{13} \right) $ to vary from $ 0 $ to $ 0.1 $ and the CP phase $ 0-2\pi $. Thus, we also take into account the uncertainties in the present values of these poorly known parameters. We will show that the uncertainties in these poorly known parameters cannot mask the effects due to neutrino decay or Lorentz-violation.

We calculate the diffuse ultra high energy neutrino fluxes 
and for specificity focus on active galactic nuclei (AGN) 
as sources for this flux. 
We demonstrate how the diffuse flux spectra 
change as a result of new physics scenarios. 
Neutrino decay, depending of the choice of the neutrino lifetime, 
results in partial-to-complete disappearance of 
the heavier neutrino mass eigenstates, leaving mainly the lightest 
mass eigenstate to be recorded in the detector\footnote{We assume that 
the lightest neutrino mass eigenstate is stable.}. 
Therefore for complete decay, 
the flavor ratios at the detector are given by 
$|U_{ei}|^2:|U_{\mu i}|^2:|U_{\tau i}|^2$, ($i=1$ or $3$) which for 
tribimaximal mixing is $  4:1:1 $ for the normal hierarchy ($i=1$), 
and $  0 : 1 : 1$ for the inverted hierarchy ($i=3$). 
However, it is possible that neutrino lifetimes are such that the decay is not complete and only occurs for the lower energy neutrinos. This would introduce an energy dependence which will change not just the ratios but also the spectral shapes.

On the other hand, the effect of Lorentz invariance violation 
is more pronounced for higher energy neutrinos. 
In particular, at higher energies a breakdown of Lorentz 
symmetry will lead to the breaking of the exact/approximate 
$\mu-\tau$ symmetry that exists for standard neutrinos. 
In fact, we will show that for values of the Lorentz invariance 
breaking parameter $a>10^{-26}$ GeV, there are almost no $ \tau $-neutrinos 
arriving at the detector above $E>10^5$ GeV, whereas the $ \nu_\mu $ 
flux is enhanced compared to its expected values. Observation of this large breaking 
of the $\mu-\tau$ symmetry at higher energies by ANITA or Auger would then be an indication 
of a possible breaking of Lorentz invariance.

In this work we also estimate the number of muon 
track and shower events in IceCube to demonstrate how this 
method involving spectral distortions 
can actually be used in the terrestrial neutrino telescopes.
We show the flavor ratios not just in terms of the diffuse fluxes 
but also in terms of the ratio of muon track to shower events. 
This event ratio is seen to have spectral distortions at the low 
energy end for the case of neutrino decay. 

The paper is organized as follows. In Section \ref{sec:agn} we briefly outline the procedure for calculating the diffuse UHE neutrino fluxes from AGN allowing standard oscillation among the flavours during the propagation of these neutrinos. In the next section, we demonstrate the effect of standard oscillations in even-ing out the shapes and magnitudes of the flavour fluxes from AGN. Section \ref{sec:decay} then shows the modification of these fluxes due to decay of the heavier neutrinos, and its effect on the number of detectable events at a large volume detector like the IceCube. We examine the effect of variation of $ \theta_{13} $ and the CP violating phase $ \delta_{CP} $ in Section \ref{sec:cpv}. We look at the effect of Lorentz-symmetry violation in Section \ref{sec:lv}, and finish with brief investigations of the effects of pseudo-Dirac neutrinos and decoherence in the last two sections.

\section{\label{sec:agn}The diffuse neutrino flux from Active Galactic Nuclei}
Active galactic nuclei are extremely distant galactic cores having very high densities and temperatures. Due to the high temperatures and the presence of strong electromagnetic fields, AGN's act as accelerators of fundamental particles, driving them to ultra-high energies ($ >1000 $ GeV). The acceleration of electrons as well as protons (or ions) by strong magnetic fields in cosmic accelerators like AGN's leads to neutrino production. Specifically, accelerated electrons lose their energy via synchrotron radiation in the magnetic field leading to emission of photons that act as targets for the accelerated protons to undergo photo-hadronic interactions. This leads to the production of mesons which are unstable and decay.  In the standard case the charged pions decay primarily contributing to neutrino production via $\pi^{\pm} \rightarrow \mu^{\pm}\nu_{\mu} $ and subsequent muon decay via $ \mu^{\pm} \rightarrow e^{\pm} \nu_{\mu} \nu_e $. This leads to a flavour flux ratio of $ \left( \nu_{e}:\nu_{\mu}:\nu_{\tau} =\right) 1:2:0 $ in the standard case. The particles finally produced as a result of this process are, thus, high energy neutrons, photons, electron pairs and neutrinos.

In this section we calculate the diffuse flux spectrum of neutrinos escaping from both optically thick AGN's, which are so called because they are opaque to neutrons and trap them, and optically thin AGN's, which are neutron-transparent, and detected at distant detectors, for instance, at IceCube \cite{Baret:2008zz}. 

To calculate the flux from optically thick sources, we use the spectra of neutrinos produced in a standard AGN source, as discussed in detail in \cite{Mannheim:1998wp}.  We then account for red-shifting in the energy dependence of the spectra appropriately. To obtain the upper bound for the diffuse AGN flux we vary the break energy $ E_b $ within the allowed range and maximally superpose all the resulting spectra. To obtain the diffuse AGN flux spectrum at earth using a standard AGN distribution across the universe, we integrate the red-shifted spectra from the individual sources over the standard AGN distribution in the universe. The resulting diffuse bound and spectrum are then normalised using the cosmic ray bounds also obtainable using a similar calculation for the cosmic ray spectrum, but here used directly from \cite{Mannheim:1998wp}.

Following \cite{Mannheim:1998wp}, we assume that the production spectra for neutrons and cosmic rays from a single AGN are given by
\begin{equation}
Q_n(E_n,L_p) \propto L_p \exp\left[\frac{-E_n}{E_{\rm max}}\right]
		\left \{ \begin{array}{ll}
                E_n^{-1}E_b^{-1} & (E_n < E_b)\\
                E_n^{-2} & (E_b < E_n)
                \end{array} \right.  ,
                \label{eq.generic_n}
\end{equation}
\begin{equation}
Q_{\rm cr}(E_p,L_p) \propto L_p \exp\left[\frac{-E_p}{E_{\rm max}}\right]
		\left \{ \begin{array}{ll}
                E_p^{-1}E_b^{-1} & (E_p < E_b)\\
                E_p^{-3}E_b      & (E_b < E_p)
                \end{array} \right.  ,
\label{eq.generic_cr}
\end{equation}
where
\begin{itemize}
\item 
$ Q_{n} $ and $ Q_{\text{cr}} $ represent the neutron and cosmic ray spectrum respectively, as a function of the neutron and proton energies $ E_n $ and $ E_p $ respectively,
\item
$ L_{p} $ represents the proton luminosity of the source,
\item
$ E_b $ is the spectrum breaking energy which can vary from $ 10^{7}$ GeV to $ 10^{10} $ GeV for optically thick AGN sources, and finally,
\item
$ E_{\text{max}} $ is the cutoff energy beyond which the spectra fall off steeply.
\end{itemize}
Using Eq.~\eqref{eq.generic_n} the generic neutrino production spectrum from AGN's can be written as
\begin{equation}
\label{generic_nu:AGN} 
	Q_{\nu_\mu}(E) \approx 83.3 Q_n(25 E) 
\end{equation}

We now need to account for red-shifting in the energies of the neutrinos propagating over cosmological distances prior to arriving at the detector. It is convenient to describe the red-shifting in terms of the dimensionless red-shift parameter $ z $, defined as
\[ \frac{\lambda}{\lambda_0}=1+z, \]
$ \lambda $ and $ \lambda_0 $ being wavelengths of a propagating signal at detector and at source respectively. In terms of $ z $ the energy of a particle at source ($ E_0 $) and at the detector ($ E $) can be related via 
\[ \frac{E_0}{E}=1+z. \]
Thus, to account for red-shifting in the energy of the neutrinos we replace the source energy $ E $ in Eq.~\eqref{generic_nu:AGN} by $ E(1+z) $. We now incorporate standard neutrino oscillations by multiplying the spectrum with the oscillation probabilities. The probability of a neutrino flavour $ \nu_\alpha $ oscillating to another $ \nu_\beta $ is given by
\ifonecol
\begin{equation}
	P_{\alpha\rightarrow\beta} = \delta_{\alpha\beta} - 4\sum_{i>j}\mathcal{R}e \left(U^*_{\alpha i}U_{\beta i}U_{\alpha j}U^*_{\beta j} \right)\sin^2 \left( \frac{\Delta m^2_{ij}L}{4E} \right)
\end{equation}
\else
\begin{equation}
\begin{split}
	P_{\alpha\rightarrow\beta} &= \delta_{\alpha\beta} \\
	                              & - 4\sum_{i>j}\mathcal{R}e \left(U^*_{\alpha i}U_{\beta i}U_{\alpha j}U^*_{\beta j} \right)\sin^2 \left( \frac{\Delta m^2_{ij}L}{4E} \right)
\end{split}
\end{equation}
\fi
However, as the distances involved are very large, oscillation only provides a $ z $-independent averaging effect over the three flavours. In all our calculations, unless otherwise mentioned\footnote{In Sec.~\ref{sec:lv} we use the tribimaximal value of $ \theta_{23} = 45^o $ that ensures perfect symmetry between $ \nu_\mu $ and $ \nu_\tau $ under standard oscillation.}, the CP violating phase $ \delta_{CP} $ is kept $ 0 $ and the 3$ \sigma $ best-fit values of the mixing angles \cite{Maltoni:2008ka} are used, i.e.,
\[
	\sin^2(\theta_{12})=0.321,\;\; \sin^2(\theta_{23})=0.47,\;\; \sin^2(\theta_{13})=0.003.
\]

The intensity at earth for an input spectrum $ Q \left[ \left(1+z \right)E,z \right] $ is given by
\begin{equation}
\label{detector_int}
I(E) \propto \int\limits_{z_{\mathrm{min}}}^{z_{\mathrm{max}}}
	\frac{(1+z)^2}{4 \pi d_L^2} \frac{\mathrm{d} V_c}{\mathrm{d}z} 
	\frac{\mathrm{d}P_{\mathrm{gal}}}{\mathrm{d}V_c}
	Q[(1+z)E,z] \, \mathrm{d}z
\end{equation}
with $ d_L $ and $ V_c $ representing the luminosity distance and co-moving volume respectively. 

To obtain the maximal bound for the diffuse flux from the optically thick sources we start with $ E_b = 10^7 $ GeV in the input spectrum $ Q $ and carry out the above integration using $ z_{min} = 0.03 $ and $ z_{max} = 6 $. The value of $ E_b $ is varied from $ 10^7 $ GeV to $ 10^{10} $ GeV and the above integration is carried out for each case. The resulting $ I_{E_b}(E) $ are then superposed to obtain the final bound. This is then normalised using the observed cosmic ray spectrum to give the upper bound of the diffuse flux for the three neutrino flavours at the detector. As may be expected, it leads to a result similar to that obtained in \cite{Mannheim:1998wp}, with, however, the results of standard oscillations incorporated. A related procedure is used for calculating the fluxes from optically-thin sources. We call this normalised upper bound of diffuse fluxes the \emph{MPR bound}, and use this as the reference flux for all our calculations. The MPR bound is a modification of the Waxman-Bahcall (W \& B) bound \cite{Waxman:1998yy}, where a uniform $ E^{-2} $ input spectrum of extragalactic cosmic rays was used to calculate the diffuse fluxes. This difference is noticeable in Fig.~\ref{fig:decay} where we have shown both these reference bounds. The resultant MPR bounds for both types of sources are shown in all the figures as unbroken gray lines.

\section{\label{sec:std-osc}Role of standard neutrino oscillations in restoring parity among flavour spectra}

As has been discussed in \cite{Learned:1994wg}, a flavour ratio of $ 1:2:0 $ at source is reduced to the democratic $ 1:1:1 $ at the detector due to oscillations. In this section we study the effect standard oscillations have on spectral shapes and magnitudes of the neutrino fluxes, as they propagate from the source to the earth. We find that AGN's being very distant sources, standard neutrino oscillations play a very significant role in restoring equality among the three flavours in terms of not only magnitudes, but also the shapes of the diffuse flux at the detector. As a result, though the flavour fluxes at some exceptional source might differ from each other in their spectral shapes significantly, neutrino oscillations during propagation ensure that such differences are largely wiped out by the time they reach the detector. In addition to spectral shapes, oscillations also have the effect of bringing widely differing magnitudes close to each other (roughly within a factor of two). This implies that if an ultra-high-energy detector at the earth detects significant difference in magnitude and energy dependence among the AGN diffuse flux of the three flavours, it must be as a consequence of non-standard physics present in the oscillation probabilities during propagation.

\begin{figure*}[ht]
	\centering
	\includegraphics[scale=0.55]{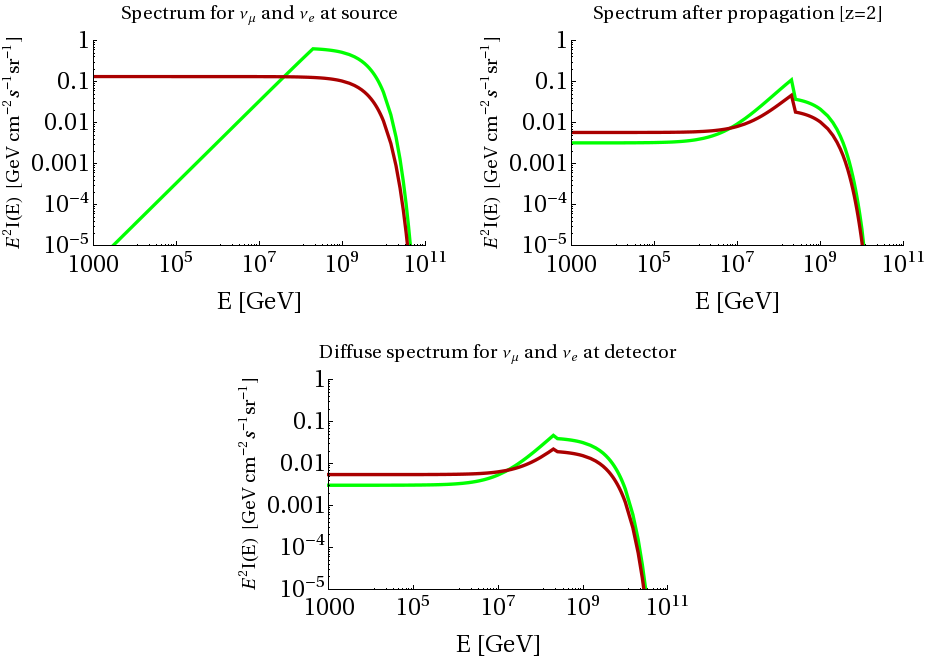}
	\caption{The even-ing out of possible spectral distortions present at source due to standard oscillations over large distances as seen for hypothetical spectra of two flavours $ \nu_\mu $ (deep-red) and $ \nu_e $ (green) from an AGN source at a redshift $ z=2 $. $ I(E) $ represents the flux spectrum for the two flavours.}
	\label{fig:osc-shape}
\end{figure*}

To demonstrate this, we assume two flavour spectra at source intentionally chosen to be widely differing and propagate them to the earth. We calculate the diffuse flux (to arbitrary normalisation) of the flavours arriving at the detector from all sources, assuming they give the same spectra for the two flavours at source. The result is shown in Fig.~\ref{fig:osc-shape}. We have checked that such a conclusion holds true in general as it does in this representative case, and demonstrates that significant differences of shape or magnitude among the diffuse flux flavours if detected must be pointers to non-standard physics playing its role during propagation.

\section{\label{sec:decay}Effect of neutrino decay}

\subsection{\label{subsec:decay_intro}Introduction to neutrino decay}

Bounds on the life-times of neutrinos are obtained primarily from observations of solar \cite{Beacom:2002cb} and atmospheric neutrinos. Observations from solar neutrinos lead to
\begin{equation}
	\frac{\tau_2}{m_2} \geq 10^{-4}\; \mathrm{s/eV}	
\end{equation}
while, if the neutrino spectrum is normal, the data on atmospheric neutrinos constrain the life-time of the heaviest neutrino
\begin{equation}
	\frac{\tau_3}{m_3} \geq 10^{-10}\; \mathrm{s/eV}.
\end{equation}

In the following, we treat the lightest neutrino as stable in view of the fact that its decay would be kinematically forbidden, and consider the decay of the heavier neutrinos to invisible daughters like sterile neutrinos, unparticle states, or Majorons. Neutrinos may decay via many possible channels. Of these, radiative two-body decay modes are strongly constrained by photon appearance searches \cite{Groom:2000in} to have very long lifetimes, as are three-body decays of the form $ \nu \rightarrow \nu \nu \bar{\nu}$ which are constrained \cite{Pakvasa:1999ta} by bounds on anomalous $ Z\nu\bar{\nu} $ couplings \cite{Bilenky:1994ma}. Decay channels of the form 
\begin{eqnarray}
\label{decay-mode-1}
\nu_i &\rightarrow& \nu_j + X \\
\label{decay-mode-2}
\nu &\rightarrow& X
\end{eqnarray}
where $ \nu_i $ represents a neutrino mass eigenstate and $ X $ represents a very light or massless invisible particle, e.g.~a Majoron, are much more weakly constrained, however and are therefore the basis of our consideration in this section. When considering decays via the channel in Eq.~\eqref{decay-mode-1} we assume that the daughter neutrino produced is significantly reduced in energy and does not contribute to the diffuse flux in the energy range relevant for our purpose ($ 1000 $ GeV to $ 10^{11} $ GeV). A detailed study of the various possible scenarios for neutrino decay is made in \cite{1126-6708-2008-07-064}.

\ifonecol
\else
\begin{figure*}
	\centering
	\includegraphics[scale=0.265]{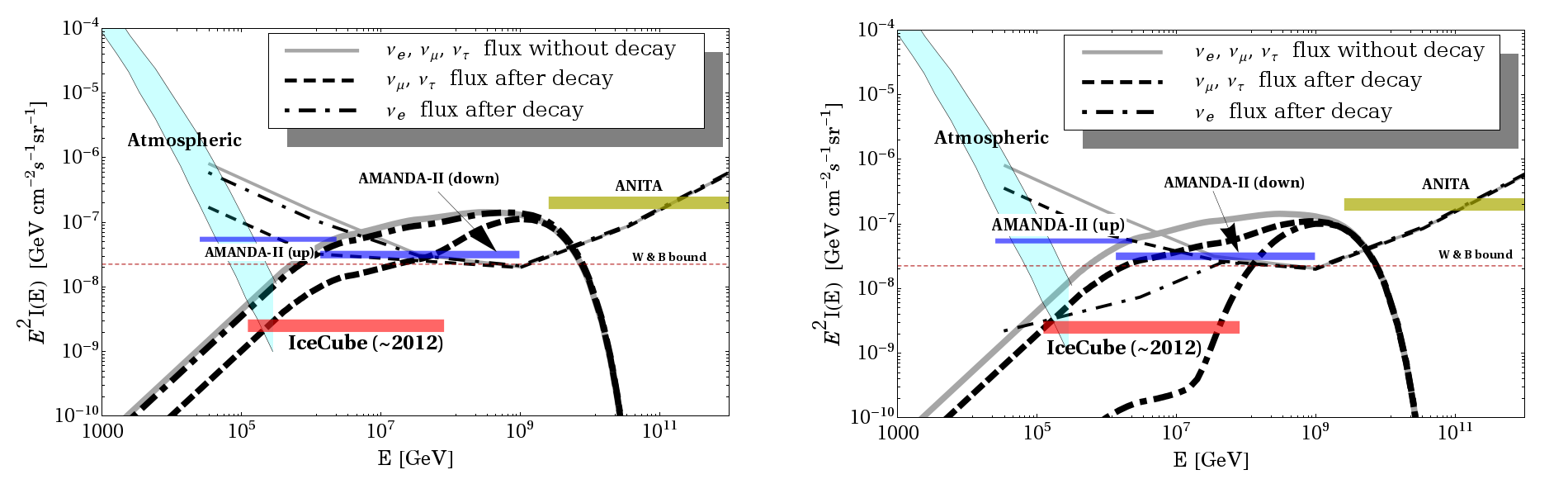}
	\caption{\label{fig:decay}Modification of MPR bound for incomplete decay with normal hierarchy (left) and inverted hierarchy (right), and life-time $ \tau_2/m_2 =  \tau_3/m_3 = 0.1 \;\mathrm{s/eV}$. The $\nu_\mu$ and $\nu_e$ fluxes shown are from optically thick (in thick) and optically thin sources (thinner). Similarly the gray lines indicate the $\nu_e,\; \nu_\mu,$ or  $\nu_\tau$ undistorted flux  modified only by neutrino oscillation, for both optically thick and thin sources. sensitivity thresholds and energy ranges of relevant experiments, \textit{viz.}, AMANDA and IceCube \cite{0004-637X-710-1-346}, and ANITA \cite{Gorham:2008yk} are indicated. $I(E)$ denotes the diffuse flux spectrum of flavours at earth, obtained as described in the text.}
\end{figure*}
\fi

Prior to proceeding, we would like to discuss cosmological observations of high precision which might be able to constrain models of decay via channels as in Eq.~\eqref{decay-mode-1} in the future. These constraints are based on the determination of the neutrino mass scale as discussed in \cite{Serpico:2007pt}, or from the cosmic microwave background as discussed in \cite{PhysRevD.72.103514}. Such observations would serve to push the lower bound of neutrino decay lifetimes by several orders of magnitude compared to those discussed here. However, these predictions are dependent upon the number of neutrinos that free-stream and assume couplings of similar nature and strength for all the species of the neutrino family. As discussed in \cite{bell:063523} and \cite{Friedland:2007vv} these assumptions must await confirmation and rely on future data. Hence, ``fast" neutrino decay scenarios are not ruled out within the scope of current theory and experiment, though they are disfavoured. Further the decay of neutrinos via Eq.~\eqref{decay-mode-2} and in the cases where the decay, both via Eq.~\eqref{decay-mode-1} and Eq.~\eqref{decay-mode-2} happen due to unparticle scenarios are not covered by such constraints and the purely phenomenological and general study of neutrino decay in the life-times discussed here would still be interesting and relevant for future neutrino detectors.

\subsection{\label{subsec:decay_cal}Effect of neutrino decay on the flavour fluxes}

A flux of neutrinos of mass $ m_i $, rest-frame lifetime $ \tau_i $, energy $ E $ propagating over a distance $L$ will undergo a depletion due to decay given (in natural units with $c=1$) by a factor of
\[
	\exp(-t/\gamma\tau)=\exp\left(-\frac{L}{E}\times \frac{m_i}{\tau_i} \right) 
\]
where $t$ is the time in the earth's (or observer's) frame and $\gamma = E/m_i$ is the Lorentz boost factor. This enters the oscillation probability and introduces a dependence on the lifetime and the energy that significantly alters the flavour spectrum. Including the decay factor, the probability of a neutrino flavour $ \nu_\alpha $ oscillating into another $ \nu_\beta $ becomes
\begin{eqnarray}
\label{decay-prob}
P_{\alpha\beta}(E) &=& \sum_{i}|U_{\beta i}|^2 |U_{\alpha i}|^2 e^{-L/\tau_i(E)},\; \alpha \neq \beta,
\end{eqnarray}
which modifies the flux at detector from a single source to
\begin{eqnarray}
\label{simple1}
\phi_{\nu_\alpha}(E) &=& \sum_{i\beta}\phi^{\rm source}_{\nu_\beta}(E)
|U_{\beta i}|^2 |U_{\alpha i}|^2 e^{-L/\tau_i(E)}.
\end{eqnarray}
We use the simplifying assumption $ \tau_2/m_2 = \tau_3/m_3 = \tau/m $ for calculations involving the normal hierarchy (\textit{i.e.} $m_3^2-m_1^2 = \Delta m_{31}^2 > 0$) and similarly, $ \tau_1/m_1 = \tau_2/m_2 = \tau/m $ for those with inverted hierarchy (\textit{i.e.} $\Delta m_{31}^2 < 0$), but our conclusions hold irrespective of this. The total flux decreases as per Eq.~\eqref{simple1}, which is expected for decays along the lines of Eq.~\eqref{decay-mode-2} and, within the limitations of the assumption made in Sec.~\ref{subsec:decay_intro}, also for Eq.~\eqref{decay-mode-1}.

The assumption of complete decay leads to (energy independent) flux changes from the expected $\nu_e^d:\nu_\mu^d:\nu_\tau^d = 1:1:1$ to significantly altered values depending on whether the neutrino mass hierarchy is normal or inverted as discussed in \cite{Beacom:2003nh}. From Fig.\ \ref{fig:decay} we note that the range of energies covered by UHE AGN fluxes spans about six to seven orders of magnitude, from about $ 10^3 $ GeV to $ 10^{10} $ GeV. For the ``no decay" case, the lowest energy neutrinos in this range should arrive relatively intact, {\it i.e.} $ L/E \simeq \tau/m\simeq 10^4 $ sec/eV. In obtaining the last number we have assumed a generic neutrino mass of $0.05$ eV and an average $L$ of 100 Mpc. On the other hand, if there is complete decay, only the highest energy neutrinos arrive intact, and one obtains \textit{i.e.} $ L/E \simeq \tau/m\leq 10^{-3} $ sec/eV. Thus, a study of the relative spectral features and differences  of flavour fluxes at earth allows us to study the unexplored range $ 10^{-3} < \tau/m < 10^4$ via  decays induced by lifetimes in this range (we have referred to this case as ``incomplete decay" in what follows).

\ifonecol
\begin{figure*}
	\centering
	\includegraphics[scale=0.25]{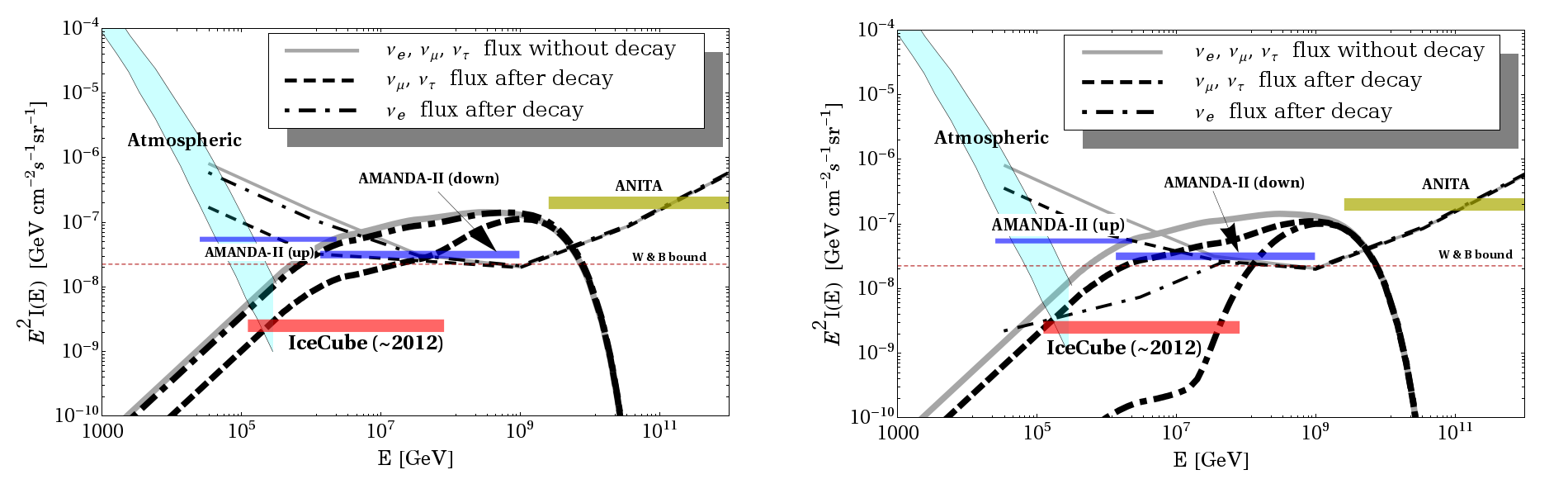}
	\caption{\label{fig:decay}Modification of MPR bound for incomplete decay with normal hierarchy (left) and inverted hierarchy (right), and life-time $ \tau_2/m_2 =  \tau_3/m_3 = 0.1 \;\mathrm{s/eV}$. The $\nu_\mu$ and $\nu_e$ fluxes shown are from optically thick (in thick) and optically thin sources (thinner). Similarly the gray lines indicate the $\nu_e,\; \nu_\mu,$ or  $\nu_\tau$ undistorted flux  modified only by neutrino oscillation, for both optically thick and thin sources. sensitivity thresholds and energy ranges of relevant experiments, \textit{viz.}, AMANDA and IceCube \cite{0004-637X-710-1-346}, and ANITA \cite{Gorham:2008yk} are indicated. $I(E)$ denotes the diffuse flux spectrum of flavours at earth, obtained as described in the text.}
\end{figure*}
\fi

To calculate the MPR-like bounds with neutrino decay we use the procedure of Sec.~\ref{sec:agn}, but replace the standard neutrino oscillation probability by $ P_{\alpha\beta} $ given in Eq.~\eqref{decay-prob} with $ E $ replaced by $ E(1+z) $ to account for red-shifting. Since, unlike standard oscillations, $ P_{\alpha\beta} $ has an energy dependence that does not just average out, the diffuse flux obtained with decay effects differ considerably from the MPR bounds in shape as well as magnitude. Fig.~\ref{fig:decay} shows the effect for both normal and inverted hierarchies with a lifetime of $ \tau_2/m_2 = \tau_3/m_3 = 0.1 $ s/eV. We note that the effect of decay in altering the diffuse flux spectrum is especially strong in the case of inverted hierarchy.

\begin{figure*}
	\centering
	\ifonecol
		\includegraphics[scale=0.295]{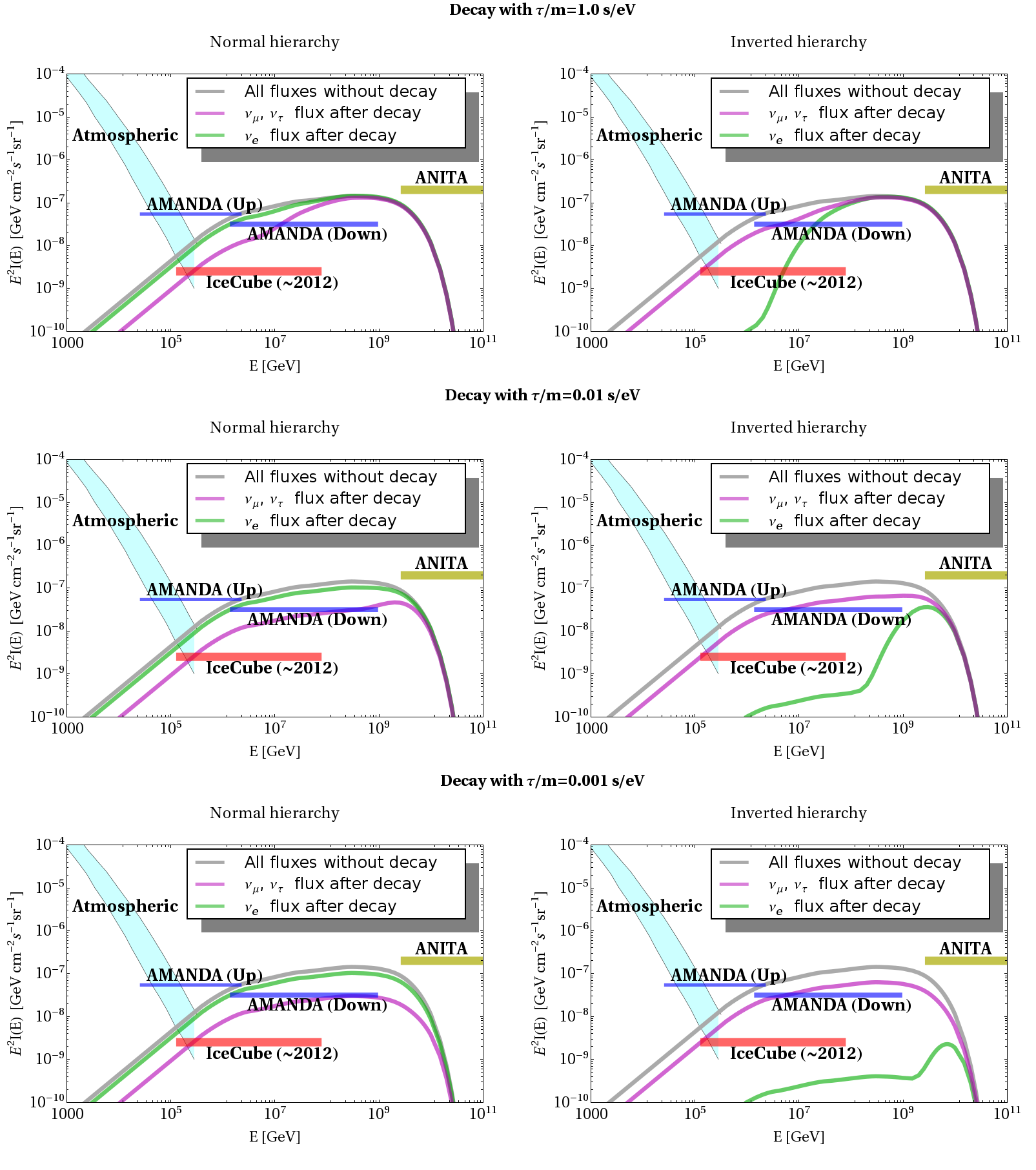}
	\else
		\includegraphics[scale=0.32]{decay_comb0}
	\fi
	\caption{\label{fig:decay2}Modification of MPR bound for incomplete decay with normal hierarchy (left) and inverted hierarchy (right), and life-times varying from $ \tau/m =  0.001 $ s/eV to $ 1.0 $ s/eV. The $\nu_\mu$ and $\nu_e$ fluxes shown are from optically thick sources. The gray lines indicate the $\nu_e,\; \nu_\mu,$ or  $\nu_\tau$ undistorted flux  modified only by neutrino oscillation. Similar effects are seen with fluxes from optically thin sources as well.}
\end{figure*}

Fig.~\ref{fig:decay2} shows how the diffuse flux spectral shapes change as the lifetimes of the two heavier mass-eigenstates are varied between $ 10^{-3} $ s/eV and $ 1 $ s/eV. From the figure it is clear that this ($ 10^{-3} $ s/eV -- $ 1 $ s/eV) is the range of life-times that can be probed by ultra-high-energy detectors looking for spectral distortions in the diffuse fluxes of the three flavours. For lifetimes above $ 1 $ s/eV the spectral shapes start to converge and become completely indistinguishable beyond $ 10^{4} $ s/eV while for those below $ 10^{-3} $ s/eV the shapes of the diffuse fluxes show no difference although their magnitudes are expectedly very different.

As is also the case for complete decays, the results are very different for the two possible  hierarchies. This is because the mass eigenstate $m_1$ contains a large proportion of $\nu_e$, whereas the state $m_3$ is, to a very large extent, just an equal mixture of $\nu_\mu$ and $\nu_\tau$ with a tiny admixture of $\nu_e$. Therefore decay in the inverted hierarchy case would lead to a disappearance of the eigenstate with high content of $ \nu_e $ and, hence, to its strong depletion against the other two flavours. In the normal hierarchy case, in comparison, the mass eigenstate with the high content of $ \nu_e $ is also the lightest, and decay of the heavier states consequently leads to a depletion of $ \nu_{\mu} $ and $ \nu_{\tau} $. Thus incomplete decay to the lowest mass eigenstate with a normal hierarchy (\textit{i.e.} $ m_1 $) would lead to considerably more shower events than anticipated with an inverted hierarchy.

While assessing the results presented here, it must be borne in mind that observation of a significant amount of $ \overline{\nu}_e $ from supernova SN1987A possibly imposes lower limits on decay lifetimes of the heavier neutrinos for the inverted hierarchy scenario that are much higher than those considered here \cite{PhysRevLett.58.1494,PhysRevLett.58.1490}. This observation, of a flux of $ \overline{\nu}_e $ roughly in keeping with standard predictions constrains its ``lifetime" $ \tau/m > 10^5 $, \textit{i.e.,} higher than what would give observable results with the methods described here. Despite the uncertainties involved with neutrino production from supernovae and the fact that the total signal from SN1987A was only a handful of events, the results for decay with inverted hierarchy must be judged keeping this in view.

\subsection{Modification of total UHE events due to decay}

The effect of decay as seen in the diffuse fluxes in Fig.~\ref{fig:decay} above must also translate to modifications in the shower and muon event rates observable at UHE detectors. In this section we demonstrate this by a sample calculation. We calculate the event-rates induced by the three flavours of high-energy cosmic neutrinos after decay using a simplified version of the procedure in  Ref.~\cite{PhysRevD.72.065019} and compare it to those predicted by standard physics.

Events at the IceCube will be classified primarily into showers and muon-tracks. Shower events are generated due to the charged current (CC) interactions of $ \nu_e $ and $ \nu_\tau $ below the energy of $1.6 $ PeV and neutral current (NC) interactions of all the three flavours. For energies greater than $ 1.6 $ PeV, CC interactions of the $ \nu_\tau $ have their own characteristic signatures in the form of double-bangs, lollipops, earth-skimming events, etc. \cite{Cowen:2007ny,Feng:2001ue}. Muon-tracks are generated due to the $ \nu_\mu $ induced CC events.

\subsubsection*{$ \mathbf{\nu_e} $ \textbf{induced events}}
In the standard model $ \nu_e $ interacts with nucleons via CC and NC interactions leading to electromagnetic and hadronic showers.

In the CC events, the shower energy is equal to the initial neutrino energy $ E_\nu $, that is,
the total energy of the two final state particles (an electron and a scattered quark). The event rate for $ \nu_e N \rightarrow e^- \chi $, with $ \chi $ being a final state quark, is given by

\begin{align}
	\text{Rate} &= \int_{E_{th}}^\infty\!\!\! \ind E_\nu \int_0^1\!\!\! \ind y\; N_A L \diff[\sigma_{CC}]{y} A \mathcal{F}\left( E_\nu \right) \\
             &= N_A V \int_{E_{th}}^\infty\!\!\! \ind E_\nu \;\sigma_{CC}(E_\nu)\mathcal{F}(E_\nu) \label{event_e_CC}
\end{align}

where
\begin{itemize}
	\item $ E_\nu $: the incident neutrino energy
	\item $ E_{th} $: detection threshold for shower events
	\item $y$: the inelasticity parameter defined as $ y \equiv 1-\frac{E_{e,\mu,\tau}}{E_\nu} $
	\item $ A, L, V $: the area, length and volume of the detector respectively
	\item $ \mathcal{F}(E_\nu) $: the flux spectrum of neutrinos in $ \text{GeV}^{-1} \text{cm}^{-2}\text{s}^{-1} $
\end{itemize}
It is assumed that the electron range is short enough such that the effective volume of the detector is identical to the instrumental volume. Using standard tabulated values of the cross-section $ \sigma_{CC} $ \cite{Gandhi:1995tf,Gandhi:1998ri} it is straightforward to evaluate the integral in Eq.~\eqref{event_e_CC} to obtain the event rate. The event rate for anti-neutrino process $ \overline{\nu}_e N \rightarrow e^+ \chi $ is calculated similarly.

For the NC events, the final state neutrino develops into missing energy, so that the rate is given by
\begin{equation}\label{event_e_NC}
	\text{Rate} = \int_{E_{th}}^\infty\!\!\! \ind E_\nu \int_{\frac{E_
{th}}{E_\nu}}^1\!\!\! \ind y\; N_A L \diff[\sigma_{NC}]{y} A \mathcal{F}\left( E_\nu \right)
\end{equation}

To simplify Eq.~\eqref{event_e_NC} we use the approximation
\begin{equation}\label{approx1}
	\diff[\sigma]{y} \approx \sigma \delta \left( y - \langle y \rangle \right)
\end{equation}
where $ \langle y \rangle $ is the mean inelasticity parameter. Thus, we have

\begin{equation}\label{event_e_NC2}
	\text{Rate} = N_A V \int_{E'_{th}}^\infty\!\!\! \ind E_\nu \;\sigma_{NC}(E_\nu)\mathcal{F}(E_\nu), 
\end{equation}
$E'_{th} $ is an effective threshold energy at which the curves defined by $ y = E_{th}/E_\nu $ and $ y=\langle y \rangle $ intersect.

\subsubsection*{\textbf{$\nu_\mu$ induced events}}
The muon track event is calculated by
\ifonecol
\begin{equation}
	\int_{E_{th}}^\infty\!\!\!\ind E_\nu\, N_A \int_0^{1-\frac{E_{th}}{E_\nu}}\!\!\! \ind y\, R \left(E_\nu (1-y),E_{th} \right) \diff[\sigma_{CC}]{y} S(E_\nu) A \mathcal{F}(E_\nu),
\end{equation}
\else
\begin{multline}
	\int_{E_{th}}^\infty\!\!\!\ind E_\nu\, N_A \int_0^{1-\frac{E_{th}}{E_\nu}}\!\!\! \ind y\, R \left(E_\nu (1-y),E_{th} \right) \\
\times \diff[\sigma_{CC}]{y} S(E_\nu) A \mathcal{F}(E_\nu),
\end{multline}
\fi
where,
\begin{equation}
	R(x,y)=\frac{1}{b} \ln \left(\frac{a+bx}{a+by}\right)
\end{equation}
with $ a=2.0 \times 10^{-3} \text{ GeV cm}^{-1} $ and $ b=3.9 \times 10^{-6} \text{  GeV cm}^{-1} $. $ S(E_\nu) $ represents the shadowing effect by the earth \cite{Gandhi:1995tf,Gandhi:1998ri}.

Approximating using Eq.~\eqref{approx1} gives 
\ifonecol
\begin{equation}
	\text{Rate}=\int_{E'_{th}}^\infty \!\!\! \ind E_\nu\, N_A R \left(E_\nu (1-\langle y \rangle),E_{th} \right) \sigma_{CC} (E_\nu) S(E_\nu) A \mathcal{F}(E_\nu)
\end{equation}
\else
\begin{multline}
	\text{Rate}=\int_{E'_{th}}^\infty \!\!\! \ind E_\nu\, N_A R \left(E_\nu (1-\langle y \rangle),E_{th} \right) \\
\times \sigma_{CC} (E_\nu) S(E_\nu) A \mathcal{F}(E_\nu)
\end{multline}
\fi
with $ E'_{th} $ being determined similarly as for the $ \nu_e $ induced events.

Using the procedure described above, we calculate the total shower and muon-track detector events (for $ \overline{\nu} + \nu $) for the inverted hierarchy scenario with a life-time of $ 1.0 $ s/eV depicted in Fig.~\ref{fig:decay2} (top-right) and compare it to the events expected from standard physics.  The results are tabulated in Table \ref{tab:decay-events} where we show event rates for UHE detectors, like the IceCube, over a 10 year period integrated over solid angle. The difference between the ratio of muon-track to shower events due to standard oscillation and that after considering neutrino decay are shown in Fig.~\ref{fig:decay3}.

\newcolumntype{.}{D{.}{.}{3,1}}
\ifonecol
\begin{table}[ht]
	\centering
	\scalebox{0.8}{
	\begin{ruledtabular}
	\begin{tabular}{c c . . c . .}
		\multicolumn{1}{c}{\textbf{Energy}} & & \multicolumn{2}{c}{\textbf{Shower}} & & \multicolumn{2}{c}{\textbf{Muon Track}} \\
		\noalign{\smallskip}\cline{3-4} \cline{6-7}\noalign{\smallskip}
		\multicolumn{1}{c}{\textbf{[GeV]}} & & \multicolumn{1}{c}{No Decay} &  \multicolumn{1}{c}{Decay} & &  \multicolumn{1}{c}{No Decay} &  \multicolumn{1}{c}{Decay} \\
		\noalign{\smallskip}\hline\noalign{\smallskip}
		$ 10^3 \; - \; 10^4 $      &  & 7       & 2      &  & 10     & 5    \\
		$ 10^4 \; - \; 10^5 $      &  & 42      & 11     &  & 96     & 42    \\
		$ 10^5 \; - \; 10^6 $      &  & 145     & 36     &  & 325    & 143   \\
		$ 10^6 \; - \; 10^7 $      &  & 129     & 24     &  & 297    & 134   \\
		$ 10^7 \; - \; 10^8 $      &  & 64      & 31     &  & 85     & 53    \\
		$ 10^8 \; - \; 10^9 $      &  & 21      & 19     &  & 16     & 14    \\
		$ 10^9 \; - \; 10^{10} $   &  & 3       & 3     &  & 1      & 1     \\
		$ 10^{10} \; - \; 10^{11} $&  & 0       & 0      &  & 0      & 0     \\
	\end{tabular}
	\end{ruledtabular}}
	\caption{\label{tab:decay-events}Total shower and muon-track detector events (for $ \overline{\nu} + \nu $) over 10 years, and integrated over solid angle for the inverted hierarchy scenario with a life-time of $ \tau/m=1.0 $ s/eV depicted in Fig.~\ref{decay2}.}
\end{table}
\else
\begin{table}[ht]
	\centering
	\begin{ruledtabular}
	\begin{tabular}{c c . . c . .}
		\multicolumn{1}{c}{\textbf{Energy}} & & \multicolumn{2}{c}{\textbf{Shower}} & & \multicolumn{2}{c}{\textbf{Muon Track}} \\
		\noalign{\smallskip}\cline{3-4} \cline{6-7}\noalign{\smallskip}
		\multicolumn{1}{c}{\textbf{[GeV]}} & & \multicolumn{1}{c}{No Decay} &  \multicolumn{1}{c}{Decay} & &  \multicolumn{1}{c}{No Decay} &  \multicolumn{1}{c}{Decay} \\
		\noalign{\smallskip}\hline\noalign{\smallskip}
		$ 10^3 \; - \; 10^4 $      &  & 7       & 2      &  & 10     & 5    \\
		$ 10^4 \; - \; 10^5 $      &  & 42      & 11     &  & 96     & 42    \\
		$ 10^5 \; - \; 10^6 $      &  & 145     & 36     &  & 325    & 143   \\
		$ 10^6 \; - \; 10^7 $      &  & 129     & 24     &  & 297    & 134   \\
		$ 10^7 \; - \; 10^8 $      &  & 64      & 31     &  & 85     & 53    \\
		$ 10^8 \; - \; 10^9 $      &  & 21      & 19     &  & 16     & 14    \\
		$ 10^9 \; - \; 10^{10} $   &  & 3       & 3     &  & 1      & 1     \\
		$ 10^{10} \; - \; 10^{11} $&  & 0       & 0      &  & 0      & 0     \\
	\end{tabular}
	\end{ruledtabular}
	\caption{\label{tab:decay-events}Total shower and muon-track detector events (for $ \overline{\nu} + \nu $) over 10 years (rounded off to whole numbers), and integrated over solid angle for the inverted hierarchy scenario with a life-time of $ \tau/m=1.0 $ s/eV depicted in Fig.~\ref{fig:decay2}.}
\end{table}
\fi

\begin{figure}
\centering
	\includegraphics[scale=0.26]{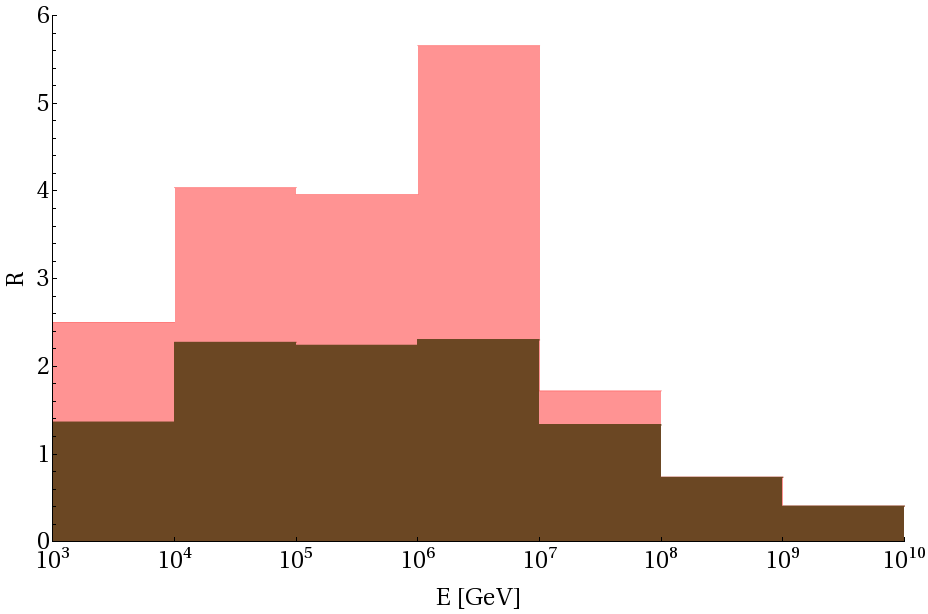}
	\caption{\label{fig:decay3}The ratio (R) of muon-track events to shower events with inverted hierarchy and life-time $ \tau/m=1.0 $ s/eV as shown in Table \ref{tab:decay-events}. The ratio expected due to standard physics is shown in brown, while the modified ratio due to the effects of decay is shown in light red. At energies greater than $10^8$ GeV, \textit{R} due to standard physics and that after considering decay become equal.}
\end{figure}

The disappearance of a majority of shower events (due to the depletion of the $\nu_e$ flux compared to that of $\nu_\mu$) at lower energies, and their reappearance at 
higher energies is a distinctive feature. It indicates the presence of new physics (like incomplete decay) as opposed to spectral distortions originating in the source, or  the appearance of a new class of sources. In the latter case, a corresponding depletion and subsequent enhancement is expected in muon events. By contrast, in the case of incomplete decay the fluxes return to the democratic ratio at higher energies where the neutrinos do not decay.

\section{\label{sec:cpv}Effect of non-zero CP violating phase and $ \theta_{13} $ variation on neutrino decay}
As described in Sec.~\ref{sec:decay} the calculation for the effect of decay of heavier neutrinos on the diffuse flux spectrum was done keeping the CP violating phase $ \delta_{CP} = 0 $ and $ \theta_{13} $ at the 3$ \sigma $ best fit value which is close to zero. In this section we look at how our conclusions are affected if we change these parameters significantly. In Sec.~\ref{subsec:cpv-1} we look at how changing $ \theta_{13} $ from $ 0 $ to the CHOOZ maximum affects the decay effected diffuse fluxes, while in Sec.~\ref{subsec:cpv-2} we examine the consequences of a non-zero CP violating phase in the same context.

\subsection{\label{subsec:cpv-1}Variation of $ \theta_{13} $}
Observations at CHOOZ \cite{Apollonio:2002gd} constrain the maximum value of $ \theta_{13} $ (90 \% confidence level) such that 
\[
	\sin^2\left( 2\theta_{13}^{\mathrm{max}} \right) = 0.10.
\]
Therefore, we have for $ \theta_{13} $ the following experimentally allowed range of values
\[
	0 \leq \theta_{13} \leq 9.1^o
\]

\begin{figure*}
	\centering
	\ifonecol
		\includegraphics[scale=0.29]{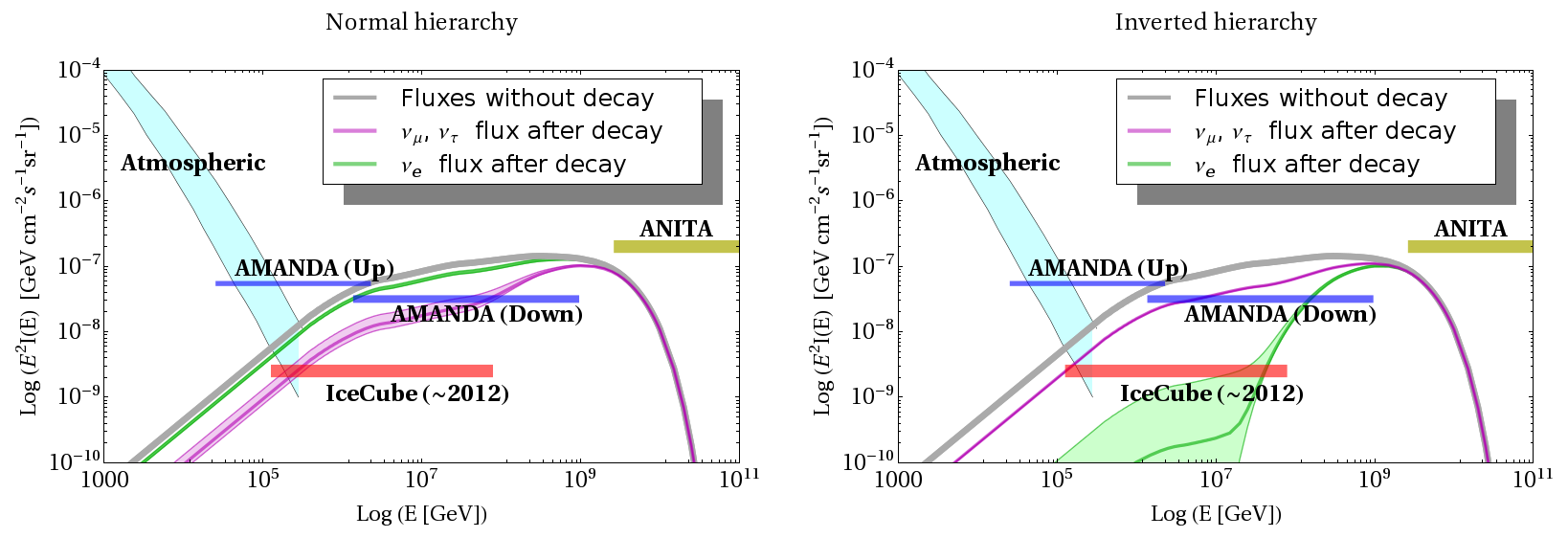}
	\else
		\includegraphics[scale=0.31]{decay_th13}
	\fi
	\caption{\label{fig:decay-th13}\textbf{Effect of variation of $ \theta_{13} $} over the complete range on decay plots obtained in Fig.~\ref{fig:decay} using optically thick sources. The shaded regions indicate the area spanned by the diffuse flux spectra as $ \theta_{13} $ varies from $ 0 $ to the CHOOZ maximum, while the thick lines represent the spectra obtained with the 3$ \sigma $ best-fit value of $ \theta_{13} $.}
\end{figure*}

We allow $ \theta_{13} $ to vary within this range and study its effect on the results of Sec.~\ref{sec:decay}. The results are represented in Fig.~\ref{fig:decay-th13}. It is clear that the effect of varying $ \theta_{13} $ is significant. However, given the strong difference in the diffuse flux spectra for inverted and normal hierarchies, variation of $ \theta_{13} $ over the entire range would not affect our qualitative conclusions in Sec.~\ref{sec:decay} regarding differentiating between the two.

\subsection{\label{subsec:cpv-2} Non-zero CP violating phase.}

The CP violation phase in the three family neutrino mixing matrix is as yet not experimentally determined. Neutrino telescopes probing ultra-high energies might be able to improve upon our present knowledge of this parameter (see \cite{Blum:2007ie} , for example). Here we look at how the presence of a non-zero CP violating phase, $ \delta_{CP} $ in the mixing matrix could affect results obtained in Sec.~\ref{sec:decay}. 

\begin{figure}[ht]
\centering
\ifonecol
	\includegraphics[scale=0.5]{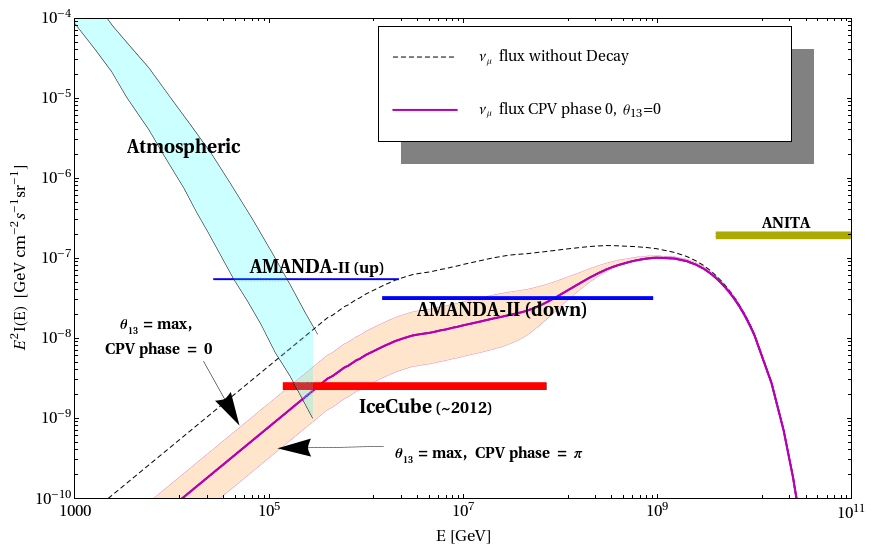}
\else
	\includegraphics[scale=0.275]{CP-1.png}
\fi
\caption{\label{fig:cpv1}\textbf{Effect of CP violation} on the diffuse flux of the $ \nu_\mu $ flavour obtained by considering decay with normal hierarchy and life-time of $ \tau/m =0.1 $ s/eV. The variation in the flux as the CP violating phase is varied between $ 0 $ -- $ \pi $ is shown as the shaded region.}
\end{figure}

\begin{figure*}
\centering
\ifonecol
\includegraphics[scale=0.29]{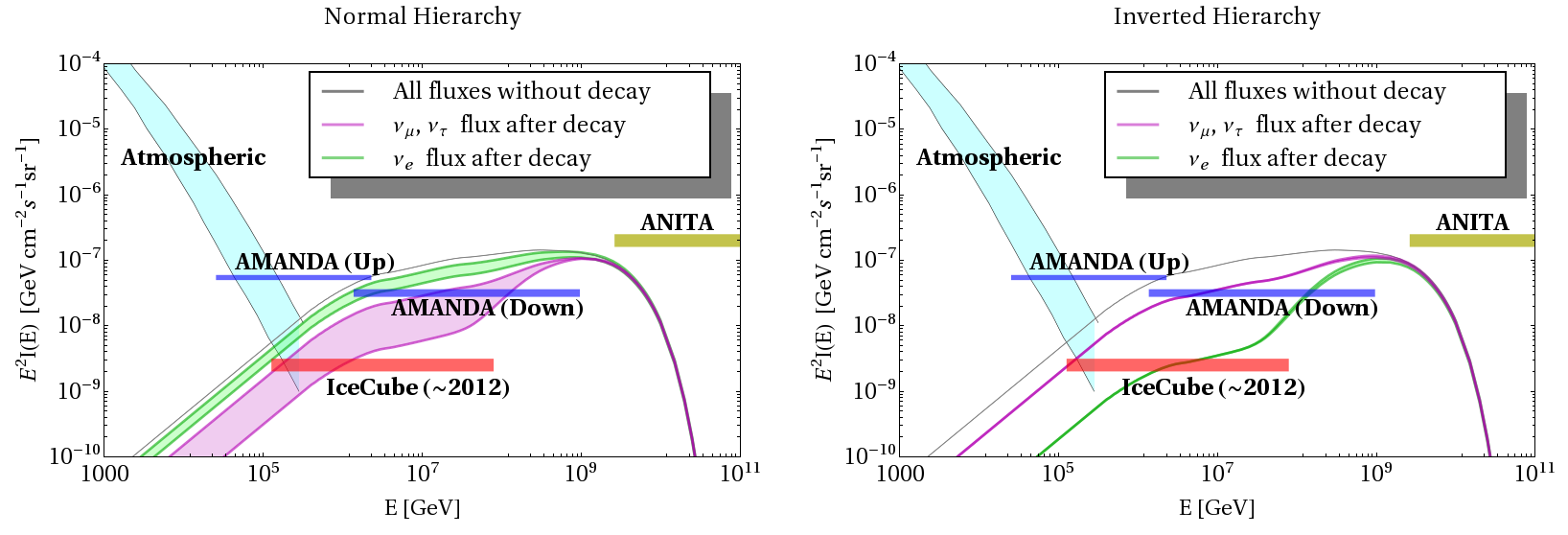}
\else
\includegraphics[scale=0.31]{CPV_comb.png}
\fi
\caption{\label{fig:cpv2}\textbf{Effect of CP violation} on fluxes affected by decay for both normal and inverted hierarchies. The shaded regions represent the span of the flux bounds when the CP violating phase is varied from $ 0 $ to $ \pi $, keeping the $ \theta_{13} $ at the CHOOZ maximum.}
\end{figure*}

$ \delta_{CP} $ enters the oscillation probability via the mixing matrix as the product $ \sin \left( \theta_{13} \right) \cdot \exp \left( \pm \imath \delta_{CP} \right) $. Therefore, a non-zero CP violating phase does not affect any of our calculations if $ \theta_{13} = 0 $ and its effect is imperceptible even when the 3$ \sigma $ best-fit value of $ \theta_{13} $ is used as is the case in Sec.~\ref{sec:decay}. For the remainder of this section we keep $ \theta_{13} $ at the CHOOZ maximum and vary the CPV phase from 0 to $ \pi $. Fig.~\ref{fig:cpv1} shows the result on the $ \nu_\mu $ flavour for decay in the case of a normal hierarchy for diffuse flux from optically thick sources. In the same way Fig.~\ref{fig:cpv2} shows the effect of a non-zero CP violating phase on decay with both the normal and inverted hierarchy. The effect of CP violation is quite small on the diffuse flux with inverted hierarchy as compared to that with normal hierarchy.

To summarise, it is clear from the discussion in Sec.~\ref{sec:decay} and \ref{sec:cpv} that future neutrino detectors capable of distinguishing between flavours should be able to probe and potentially provide stronger bounds on decay lifetimes of heavier neutrinos. If the neutrinos decay with a lifetime within the ranges discussed here, then they would also be able to distinguish between the two hierarchies due to the strongly different diffuse flux spectra the two hierarchies lead to for the flavours $ \nu_e $ and $ \nu_\mu $, notwithstanding the effect of a non-zero CP violating phase or the uncertainty over the value of $ \theta_{13} $.

\section{\label{sec:lv}Effect of Lorentz symmetry violation}

Low energy phenomenology can be affected by Lorentz symmetry violating effects originating at very high energies. Typically such effects originate at energies close to the Planck scale. They may appear in certain theories which are low energy limits of string theory \cite{Madore:2000en,Kostelecky:2003fs}, or could possibly signal the breakdown of the CPT theorem \cite{Greenberg:2002uu}. Additionally, if quantum gravity demands a fundamental length scale, leading to a breakdown of special relativity, or loop quantum gravity \cite{Rovelli:1994ge, Gambini:1998it, Alfaro:1999wd, Thiemann:2001yy, AmelinoCamelia:2003xp, Freidel:2003sp} leads to discrete space-time, one expects tiny LV effects to percolate to lower energies. UHE neutrinos, with their high
energies and long oscillation baselines present a unique opportunity for testing these theories. Their effects in the context of flavour flux ratios have been discussed in \cite{Hooper:2005jp}. They may arise, for example, due to a vector or tensor field forming a condensate and getting a vacuum expectation value, thereafter behaving like a background field. The effective contribution of such background fields can then be handled in the low energy theory using standard model extensions \cite{Kostelecky:2003fs}. It has been shown \cite{Greenberg:2002uu} that although CPT symmetry violation implies Lorentz violation, Lorentz violation does not necessarily require or imply the violation of CPT symmetry. In this section we focus on the modification of the propagation of neutrinos due to Lorentz symmetry violating effects along the lines discussed in Ref.~\cite{Kostelecky:2003cr}. Since the effects of Lorentz-violation and CPT violation are understandably tiny at low energies, it is difficult to explore their phenomenological signatures using low energy probes, in colliders for example. Since they originate in extremely energetic cosmological accelerators and propagate over cosmic distances, ultra-high energy neutrinos provide the perfect laboratory for constraining and, possibly, determining Lorentz-violating parameters.

\subsection{\label{subsec:lv-1}Modification of neutrino transition probabilities due to LV effects}
As an example, we will study, for the simplification that it provides, a two-flavour scenario with massive neutrinos and consider the modification of the transition probability from one flavour to the other by Lorentz-violation due to an effective standard model extension. Our focus is on LV from off-diagonal terms in the effective hamiltonian describing the propagation of the neutrinos \cite{Hooper:2005jp}.

We consider an effective Hamiltonian describing neutrino propagation
\begin{equation}
	H_{\alpha\beta}^{\mathrm{eff}}=\mid \vec{p}\mid \delta_{\alpha\beta} + \frac{1}{2\mid \vec{p}\mid} \left[\tilde{m}^2 + 2 \left( a^{\mu}p_{\mu} \right) \right]_{\alpha\beta}
\end{equation}
where $ \tilde{m} $ is related to the neutrino mass and $ a $ is a real CPT and Lorentz violating parameter. In the two neutrino mass basis this gives
\begin{equation}
	H_{\mathrm{eff}}=\begin{pmatrix}
			\frac{m_1^2}{2E}	& a 			\\
			a                	& \frac{m_1^2}{2E} 	\\
		\end{pmatrix}.
\end{equation}

With the mixing angle between the two flavours $ \theta_{23} = \pi/4 $, this modifies the probability of transition from one flavour to another during propagation to
\begin{equation}\label{probability-lv}
	P\left[ \nu_\mu \rightarrow \nu_\tau \right] = \frac{1}{4}\left( 1 - \frac{a^2}{\Omega^2} - \frac{\omega^2}{\Omega^2}\cos \left( 2\Omega L \right) \right)
\end{equation}
where $ \omega=\frac{\Delta m^2}{4E} $ and $ \Omega = \sqrt{\omega^2 + a^2} $.

\subsection{\label{subsec:lv-2}Effect of Lorentz violation on neutrino flavour fluxes}
To calculate the diffuse fluxes of the two neutrino flavours we use Eq.~\eqref{probability-lv} instead of the standard oscillation probability and integrate over the red-shift $ z $. The probability above contributes a $z $ dependent term through its dependence on energy. Further the $ \cos \left( 2\Omega L \right) $ term averages out and consequently does not contribute.

\begin{figure*}
\centering
\ifonecol
\includegraphics[scale=0.29]{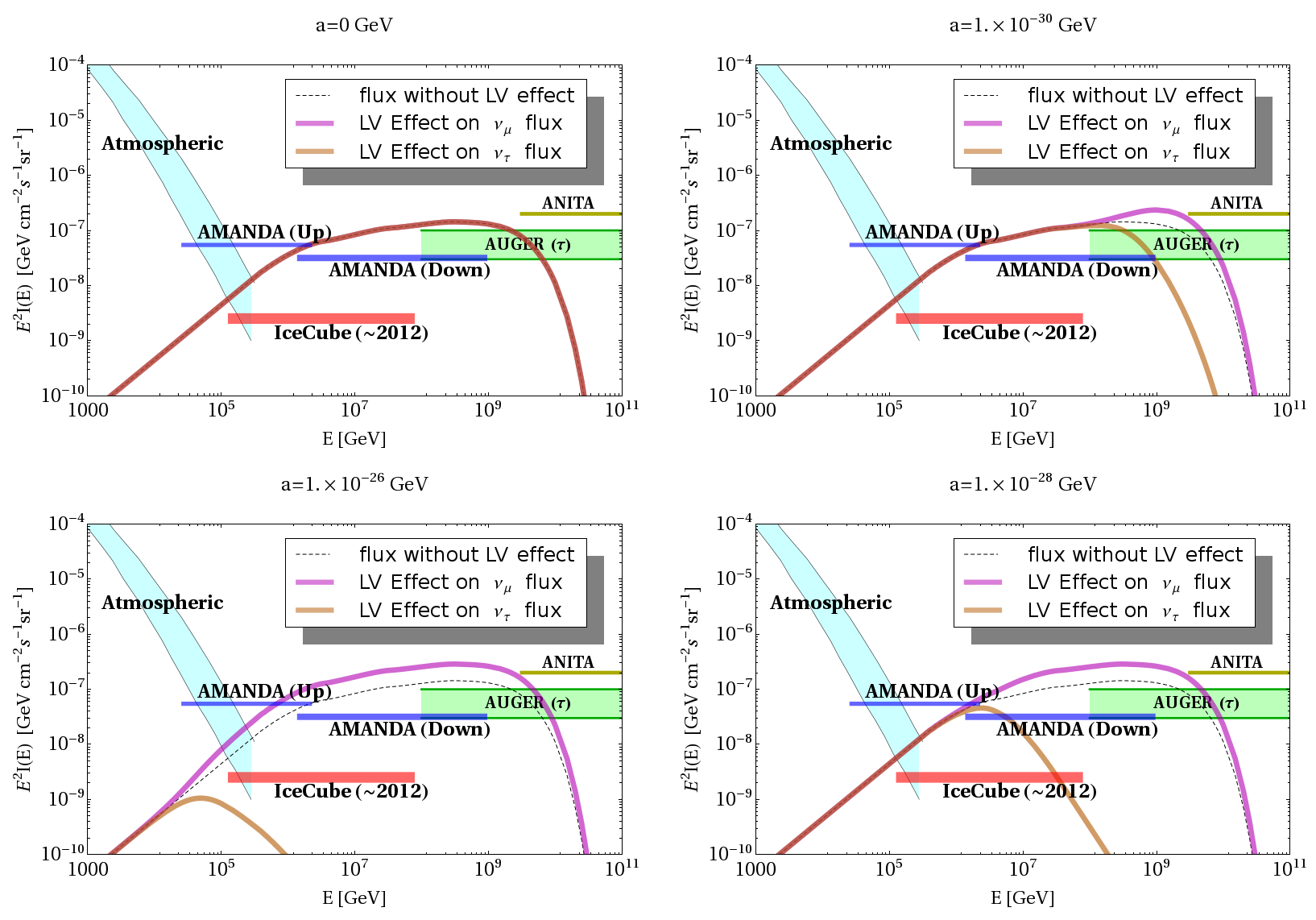}
\else
\includegraphics[scale=0.31]{LV-all.png}
\fi
\caption{\label{fig:lv1}\textbf{Effect of Lorentz violation} on the $ \nu_{\mu}-\nu_{\tau} $ diffuse flux with various values of the lorentz violating parameter $ a $ (in GeV). Clockwise from top-left (i) $ a = 0 $, (ii) $ a = 10^{-30} $, (iii) $ a = 10^{-28} $, (iv) $ a = 10^{-26} $. The plots show how an increase in the LV parameter results in depletion of the $ \nu_{\tau} $ flux at progressively lower energies. For the Auger experiment, sensitivities for $ \nu_\tau $ detection using the most pessimistic systematics (top line) and the most optimistic systematics (bottom line) are indicated \cite{PhysRevD.79.102001}.}
\end{figure*}

The results of including Lorentz violation in the propagation phenomenology of neutrinos are shown in Fig.~\ref{fig:lv1}.
It is clear from these plots that the strong departure of diffuse spectral shapes of $ \nu_\mu $ and $ \nu_\tau $ from the symmetry expected under standard oscillation phenomenology with $ \theta_{23}=45^o $ is a unique signature of Lorentz-violation. This would lead to a significant decrease in the signature $ \nu_\tau $ events at high energies, like ``double-bang", ``lollipop" and ``earth-skimming" events as compared to muon-track events. Differences in shape between the two flavours can be seen for $ a < 10^{-30} $ GeV. We have used the case where $ a $ is independent of energy, however if the parameter $ a \propto E^n $ the results would be qualitatively similar to that obtained here but involve significantly different ranges of values for the parameter as expected.

\subsection{Detectability of Lorentz-violation}
Unlike in neutrino decay, the effect of Lorentz violation is seen in the deviation of the flux spectra of both the $\nu_\mu $ and, more strikingly, the $ \nu_\tau $ flavour, from the standard fluxes toward the higher end of the spectrum. This makes it especially interesting for probe by detectors, such as ANITA and the Pierre Auger Observatory \cite{PhysRevD.79.102001,Abraham:2010mj} having sensitivity to $ \nu_\tau $ in the energy range $ 10^{8} - 10^{11} $ GeV. While Auger can separate out the $ \nu_\tau $ events, ANITA detects the sum of all three flavours. As is clear from the experimental thresholds shown in Fig.~\ref{fig:lv1}, should even tiny Lorentz-violation effects exist, both these experiments will, in principle, be able to detect it via lack of characteristic $ \tau $ events expected at these energies from standard physics. As they collect more data in the future, expectedly bringing the corresponding thresholds down, the ability of such experiments to detect tiny LV effects will be gradually enhanced.

\section{\label{sec:pseudo-Dirac}Pseudo-Dirac neutrinos}

Masses for neutrinos can be generated by extending the Standard model to include right-handed sterile neutrinos to the particle spectrum. The generic mass term for neutrinos becomes
\begin{equation}
	\mathfrak{L}=-\frac{1}{2}\overline{\Psi^C}M\Psi + h.c.,
\end{equation}
where considering 3 right-handed neutrinos in the spectrum
\[
	\Psi=\left( \nu_{eL},\;\nu_{\mu L},\;\nu_{\tau L},\;\left( \nu_{1R} \right)^C,\;\left( \nu_{2R} \right)^C,\;\left( \nu_{3R} \right)^C \right),
\]
and $ \nu^C = \mathcal{C}\overline{\nu}^T $, $ \mathcal{C} $ being the charge conjugation operator.

The mass matrix $ M $ is of the form
\begin{equation}
	M = \begin{pmatrix}
		m_L & m_D^T \\
		m_D & m^*_R
	\end{pmatrix},
\end{equation}
and for $ m_L=m_R=0 $ reduces to neutrino states with Dirac mass. In this case the six neutrinos decompose into three active-sterile pairs of neutrinos degenerate in mass with maximal mixing angle $ \theta=\pi/4 $ for each pair. Due to the mass degeneracy within the neutrinos in such a pair, an active neutrino cannot oscillate into a sterile neutrino from the same pair.

Instead, neutrinos may be pseudo-Dirac states \cite{Beacom:2003eu} where $ m_L $ and $ m_R $ are tiny but non-zero, i.e.~$ m_L,\,m_R \ll m_D $. This lifts the degeneracy in mass within an active-sterile pair, and gives a mixing angle $ \theta \approx \pi/4 $ between its members. The result of the lifting of this degeneracy is to enable oscillation among species that was not possible in the pure Dirac neutrino case.

The presence of non-zero $ m_L,\,m_R $ changes the probability of transition of one active state to another during propagation. The expression for the probability for neutrinos propagating over cosmological distances (after various phase factors involving terms like $ \Delta m^2_\odot/L $ average out) is \cite{Beacom:2003eu}
\begin{equation}\label{prob-pseudo}
	P_{\alpha\beta}=\sum_{j=1}^3 \mid U_{\alpha j} \mid^2 \, \mid U_{\beta j} \mid^2 \cos^2\left( \frac{\Delta m_j^2 L}{4E_\nu} \right),
\end{equation}
where $ \Delta m^2_j = \left( m_j^+ \right)^2 -\left( m_j^- \right)^2 $ is the mass squared difference between the active and sterile states in the $ j^{\mathrm{th}} $ pair. 

\begin{figure}
	\centering
	\ifonecol
		\includegraphics[scale=0.5]{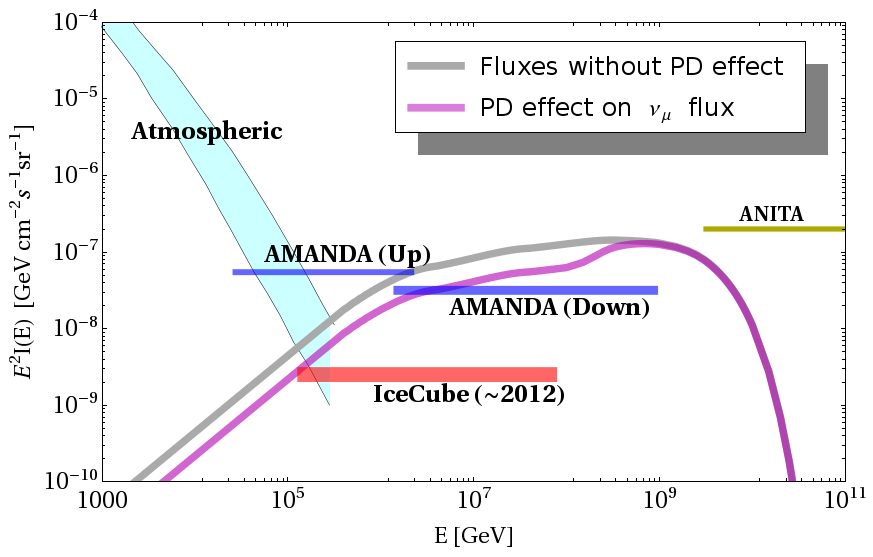}
	\else
		\includegraphics[scale=0.28]{pseudoD_14.png}
	\fi
	\caption{\label{fig:pseudo1}\textbf{Effect of pseudo-Dirac (PD) neutrinos} on the $ \nu_{\mu} $ diffuse flux with $ \Delta m^2 = 10^{-14}\, \mathrm{eV}^2 $.}

\end{figure}

There has been a recent study \cite{Esmaili:2009fk} that explores the pseudo-Dirac scenario at neutrino telescopes using the ratio of shower to muon-track events. Here, we look at distortion of spectral shape from the standard diffuse flux due to the modification of the oscillation probability to Eq.~\eqref{prob-pseudo}. We use Eq.~\eqref{prob-pseudo} instead of the standard oscillation probability, otherwise following the same procedure used to derive the standard MPR flux (the base flux in our plots). The results are shown in Fig.~\ref{fig:pseudo1} which shows a decrease in the affected flux at lower energies and rise at the higher end of the spectrum to merge with the standard flux. However, the decrease is only to about half the base flux, and the rise at higher energies is not steep. Therefore, it would be very difficult to detect such an effect in future detector experiments.

\section{\label{sec:decoherence}Effect of decoherence during neutrino propagation}
Quantum decoherence arises at the Planck scale in theories where CPT invariance is broken independently of Lorentz symmetry due to loss of unitarity and serves to modify the time evolution of the density matrix \cite{Hooper:2005jp, Hooper:2004xr}. Though not expected in a majority of string theories, a certain class of string theories called noncritical string theories may allow for decoherence.

In the context of neutrino oscillation, decoherence serves to modify the transition probabilities among the three flavours. While a general treatment discussing how this happens for the three family case is complicated, we work under the simplifying conditions assumed in \cite[see Sec IV.B]{Hooper:2005jp} to arrive at the transition probability
\ifonecol
\begin{multline}
		P \left[\nu_p \rightarrow \nu_q \right]  = \frac{1}{3} + \frac{1}{6} e^{-2\delta L}\left[ 3\left( U^2_{p1} - U^2_{p2} \right)\left( U^2_{q1} - U^2_{q2} \right) \right. \\
		  \quad + \left. \left( U^2_{p1} + U^2_{p2} - 2 U^2_{p3}\right)\left( U^2_{q1} + U^2_{q2} - 2 U^2_{q3}\right) \right],
\end{multline}
\else
\begin{equation}
\begin{split}
		P \left[\nu_p \rightarrow \nu_q \right] &= \frac{1}{3} + \frac{1}{6} e^{-2\delta L}\left[ 3\left( U^2_{p1} - U^2_{p2} \right)\left( U^2_{q1} - U^2_{q2} \right) \right. \\
		  & + \left. \left( U^2_{p1} + U^2_{p2} - 2 U^2_{p3}\right)\left( U^2_{q1} + U^2_{q2} - 2 U^2_{q3}\right) \right],
\end{split}
\end{equation}
\fi
where $ \delta $ is the only decoherence parameter. This leads to a flavour composition at the detector given by
\ifonecol
\begin{subequations}\label{decoh_flavour}
\begin{align}
	R_{\nu_e} &= P\left[ \nu_e \rightarrow \nu_e \right]\frac{\Phi_{\nu_e}}{\Phi_{\mathrm{\textsc{tot}}}} + P\left[ \nu_\mu \rightarrow \nu_e \right]\frac{\Phi_{\nu_\mu}}{\Phi_{\mathrm{\textsc{tot}}}} + P\left[ \nu_\tau \rightarrow \nu_e \right]\frac{\Phi_{\nu_\tau}}{\Phi_{\mathrm{\textsc{tot}}}}\,, \\
	R_{\nu_\mu} &= P\left[ \nu_e \rightarrow \nu_\mu \right]\frac{\Phi_{\nu_e}}{\Phi_{\mathrm{\textsc{tot}}}} + P\left[ \nu_\mu \rightarrow \nu_\mu \right]\frac{\Phi_{\nu_\mu}}{\Phi_{\mathrm{\textsc{tot}}}} + P\left[ \nu_\tau \rightarrow \nu_\mu \right]\frac{\Phi_{\nu_\tau}}{\Phi_{\mathrm{\textsc{tot}}}}\,, \\
	R_{\nu_\tau} &= P\left[ \nu_e \rightarrow \nu_\tau \right]\frac{\Phi_{\nu_e}}{\Phi_{\mathrm{\textsc{tot}}}} + P\left[ \nu_\mu \rightarrow \nu_\tau \right]\frac{\Phi_{\nu_\mu}}{\Phi_{\mathrm{\textsc{tot}}}} + P\left[ \nu_\tau \rightarrow \nu_\tau \right]\frac{\Phi_{\nu_\tau}}{\Phi_{\mathrm{\textsc{tot}}}}\,,
\end{align}
\end{subequations}
\else
\begin{subequations}\label{decoh_flavour}
\begin{equation}
	\begin{split}
	R_{\nu_e} &= P\left[ \nu_e \rightarrow \nu_e \right]\frac{\Phi_{\nu_e}}{\Phi_{\mathrm{\textsc{tot}}}} + P\left[ \nu_\mu \rightarrow \nu_e \right]\frac{\Phi_{\nu_\mu}}{\Phi_{\mathrm{\textsc{tot}}}} \\
	&\quad + P\left[ \nu_\tau \rightarrow \nu_e \right]\frac{\Phi_{\nu_\tau}}{\Phi_{\mathrm{\textsc{tot}}}}\,,
	\end{split}
\end{equation}
\begin{equation}
	\begin{split}
	R_{\nu_\mu} &= P\left[ \nu_e \rightarrow \nu_\mu \right]\frac{\Phi_{\nu_e}}{\Phi_{\mathrm{\textsc{tot}}}} + P\left[ \nu_\mu \rightarrow \nu_\mu \right]\frac{\Phi_{\nu_\mu}}{\Phi_{\mathrm{\textsc{tot}}}} \\
	&\quad + P\left[ \nu_\tau \rightarrow \nu_\mu \right]\frac{\Phi_{\nu_\tau}}{\Phi_{\mathrm{\textsc{tot}}}}\,,
	\end{split}
\end{equation}
\begin{equation}
	\begin{split}
	R_{\nu_\tau} &= P\left[ \nu_e \rightarrow \nu_\tau \right]\frac{\Phi_{\nu_e}}{\Phi_{\mathrm{\textsc{tot}}}} + P\left[ \nu_\mu \rightarrow \nu_\tau \right]\frac{\Phi_{\nu_\mu}}{\Phi_{\mathrm{\textsc{tot}}}} \\
	&\quad + P\left[ \nu_\tau \rightarrow \nu_\tau \right]\frac{\Phi_{\nu_\tau}}{\Phi_{\mathrm{\textsc{tot}}}}\,,
	\end{split}
\end{equation}
\end{subequations}
\fi
where $ \Phi_{e}/\Phi_{\mathrm{\textsc{tot}}} $, etc.~are flux composition ratios at source.

\ifonecol
\begin{figure}[ht]
	\centering
		\includegraphics[scale=0.285]{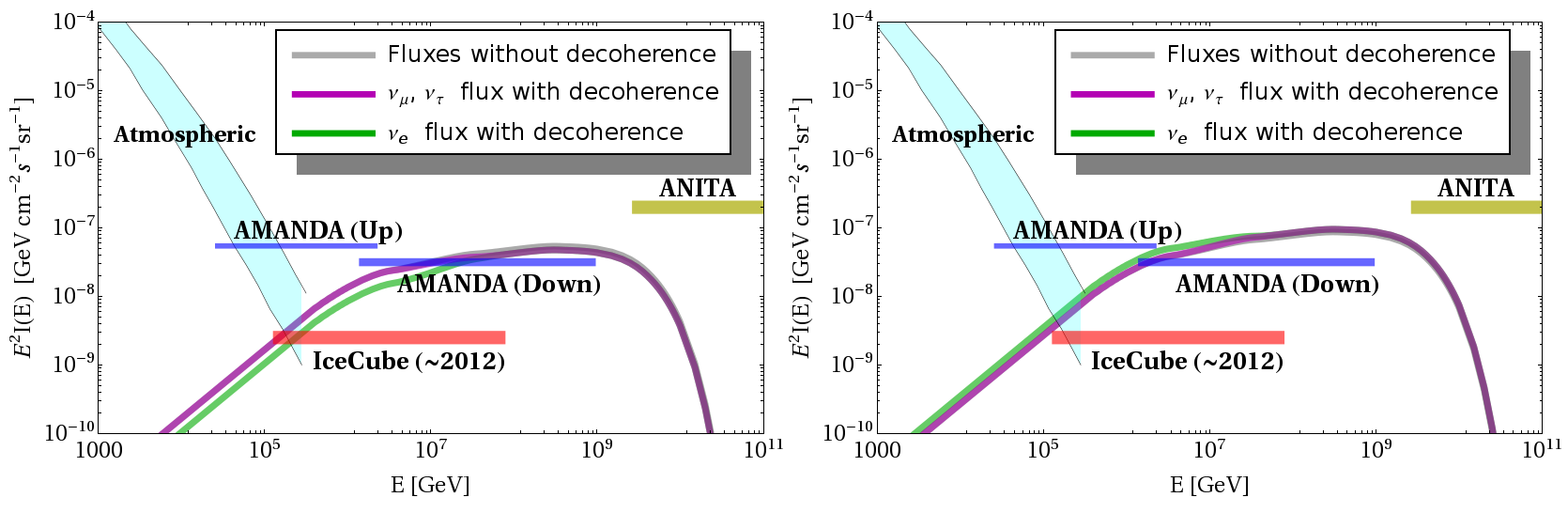}
	\caption{\label{fig:decoherence-1}\textbf{Effect of decoherence} on the diffuse flux with the parameter $ \delta=\alpha E^2 $ and $ \alpha=10^{-40} \text{ GeV}^{-1} $. A base flux composition of $ 0:1:0 $ corresponding to $ \overline{\nu} $ (\textit{left}) and $ 1:1:0 \text{ corresponding to } \nu $ (\textit{right}) from pion decay is used for the calculation. It is clear from the figure that (anti-)neutrinos from pion decay are not useful probes for decoherence.}
\end{figure}
\else
\begin{figure*}
	\centering
		\includegraphics[scale=0.305]{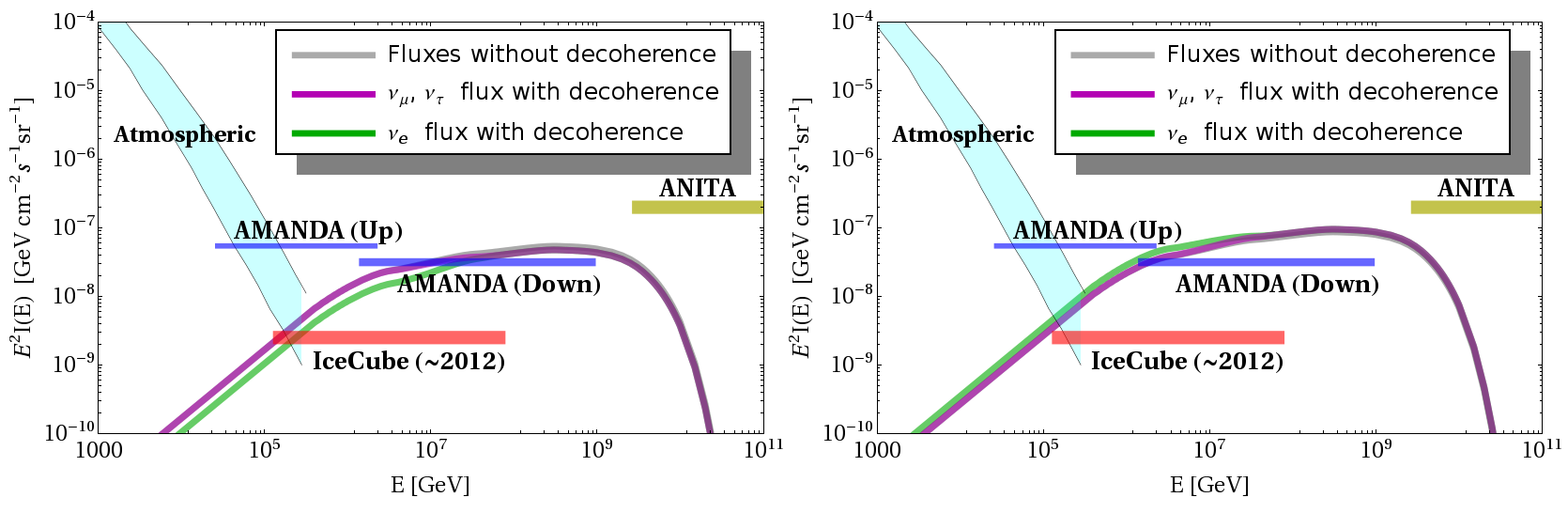}
	\caption{\label{fig:decoherence-1}\textbf{Effect of decoherence} on the diffuse flux with the parameter $ \delta=\alpha E^2 $ and $ \alpha=10^{-40} \text{ GeV}^{-1} $. A base flux composition of $ 0:1:0 $ corresponding to $ \overline{\nu} $ (\textit{left}) and $ 1:1:0 \text{ corresponding to } \nu $ (\textit{right}) from pion decay is used for the calculation. It is clear from the figure that (anti-)neutrinos from pion decay are not useful probes for decoherence.}
\end{figure*}
\fi

We use the flavour ratios given by Eq.~\eqref{decoh_flavour} to calculate the diffuse flux spectra of each flavour arriving at the detector. The effect of decoherence is to bring the flavour fluxes close to the ratio $  1:1:1 $. If we use the standard flux from AGN's ($1:2:0$ at source) then standard neutrino oscillation already brings the ratio to the above value as discussed in Sec.~\ref{sec:std-osc} and this makes it difficult to distinguish between the effects of decoherence and standard oscillation. However, if we have detection capabilities that can distinguish between neutrinos and anti-neutrinos, it might be worth investigating decoherence using the differences in flavour spectral shapes. As discussed earlier pion decays in the source via $\pi^{+} \rightarrow \nu_{\mu}\mu^{+}$ and subsequently, $\mu^{+} \rightarrow e^{+} \overline{\nu}_{\mu} \nu_e  $ contribute to a flavour spectral ratio of $ 1:1:0 $ for $ \nu $ and $ 0:1:0 $ for $ \overline{\nu} $. Due to standard oscillation these flavour ratios are reduced to $ 0.78:0.61:0.61 $ and $ 0.22:0.39:0.39 $ at the detector respectively. Since the effect of decoherence is to reduce the flavour ratios to $ 1:1:1 $ irrespective of ratios at source, the transition from the flux due to dominance of standard oscillation to that due to dominance of decoherence might happen within the energy range relevant for our purposes, for a certain range of values of the decoherence parameter. However the effect is almost invisible even if $ \nu $ and $ \overline{\nu} $ fluxes are used as probes, the reason being that the fluxes ratios at detector due to standard oscillation for both (\textit{i.e.}, $ 0.78:0.61:0.61 $ and $ 0.22:0.39:0.39 $ respectively) are already quite close to the $ 1:1:1 $ that decoherence would result in. Effective probe for decoherence are high energy neutrinos from neutron decay, for instance, which gives a flux ratio of $ 1:0:0 $ at source \cite{PhysRevD.72.065019}, and not neutrinos from pion decay. The results for $ \overline{\nu} $ and $ \nu $ with a particular choice of the decoherence parameter is shown in Fig.~\ref{fig:decoherence-1}. For our calculation, we have chosen the parameter $ \delta \propto E^2 $ which is expected within the context of string theories\footnote{The choice of $ \delta \propto E^2 $ also violates Lorentz symmetry which introduces weaker secondary effects not taken into account here.}. Upper limits on such a parameter are got from the Super-Kamiokande as $ \sim 10^{-10} \text{ GeV} $.

\section{\label{sec:conclusion}Conclusions}

In this article we have discussed the effects of several exotic, non-standard physics on the diffuse fluxes of the three neutrino flavours, using neutrino fluxes from AGN's as an example. We have assumed a standard neutrino flux at source with the flavour ratio thereof being $ 1:2:0 $ and shown that due to standard oscillations in vacuum during the propagation of these neutrinos across cosmological distances the fluxes are evened out to the democratic value of $ 1:1:1 $, and that even for non-standard fluxes at source the fluxes at the detector are still close to each other in magnitude and their spectral shapes are very similar.

Non-standard physics serves to destroy this equality among the three flavours and this serves as a potential probe for the underlying nature of the physics involved. To demonstrate this we first looked at how the decay of the heavier of the neutrinos affects the standard MPR diffuse flux bounds in the case of both normal and inverted hierarchies. We found that decay life-times of magnitudes several orders above those currently understood from experiment induce detectable changes in spectral shapes of the three diffuse fluxes, both against the standard flux, and among each other. Since the effects are strikingly different for the two hierarchies, it would also be possible to search for the hierarchy in case the heavier neutrinos do decay with life-times in the range $ 10^{-3} $ s/eV -- $ 10^4 $ s/eV, as discussed here. We have also shown that the effects remain significant despite variation on the unknown parameters $ \theta_{13} $ and $ \delta_{CP} $ and probing neutrino decay within the life-times explored here should be possible despite our limited knowledge about these parameters.

Tiny effects of Lorentz symmetry violation in the low energy theory arising due to the effect of some Planck scale physics can also be probed using ultra-high energy neutrinos. Taking the simplifying case of two neutrino flavours $ \nu_\mu $ and $ \nu_\tau $ we have described the effect of Lorentz violating parameters on transition probabilities between them during propagation and inferred that it leads to a strong decrease in the $ \nu_\tau $ flux as compared to the $ \nu_\mu $ flux. This breaks the $ \nu_\mu - \nu_\tau $ symmetry that is a feature of all standard model and most beyond standard-model scenarios, and thus provides us with a distinctive signature for LV. It translates to a corresponding decrease in the signature $ \nu_\tau $ events at high energies. While a simplifying case of two flavours and involving just the one Lorentz-violating parameter was dealt with here, the conclusions are true more generally. Detection of a sharp decrease in $ \tau $ events in future detectors like Auger and ANITA will be an indicator of the extent of Lorentz violation in low energies. Conversely, the failure to detect such a dip could be used to put bounds on the LV parameters.

Further, we have discussed the effect of decoherence and the existence of pseudo-Dirac neutrino states on the diffuse fluxes of the three flavours. While not as striking as the effects of neutrino decay or LV, the existence of pseudo-Dirac states affects distortions in the spectral shape of the standard flux at the lower end of the spectrum. On the contrary, decoherence shows almost no distortion on the fluxes. A probe of decoherence requires that we distinguish between neutrinos and anti-neutrinos since, irrespective of the flux ratio at source, it tries to bring the flux ratio to $ 1:1:1 $ at the detector, same as what standard oscillation does to the standard flux of $ 1:2:0 $. Even so, the effect of decoherence, seen at higher energies, is not significant and cannot, in all probability, be experimentally distinguished.

It is clear that future ultra-high energy neutrino detectors with strong flavour detection capabilities and excellent energy resolution will allow us to probe the validity of non-standard physical phenomena over large ranges of the involved parameters. While differences in spectra among the flavours arise due to the selectivity of non-standard physics with regard to the three families, strong distortion of spectral shape of the fluxes as compared to the standard flux expected at the detector arises due to the non-trivial energy dependence of transition probabilities in new physics. To detect or, potentially, constrain new physics it is necessary to carry out experiments that combine searches of both kinds. While understandably challenging, it will certainly be worthwhile carrying out detection experiments along these lines given the fundamental nature of physics that will be brought under the scanner.

\textbf{\textit{Acknowledgment:}} RG would like to thank Nicole Bell, Maury Goodman, Francis Halzen, Chris Quigg, Georg Raffelt, Subir Sarkar and Alexei Smirnov  for helpful discussions, and the theory groups at Fermilab and Brookhaven National Lab for hospitality while this work was in progress. This work has been supported by the Neutrino Project under the XI Plan of Harish-Chandra Research Institute.

\bibliography{long_1}{}
\end{document}